\documentclass[prc,showpacs,nofootinbib]{revtex4}
\usepackage{feynmp,epsfig,graphics,color}
\def\sign{{\rm sign}\,}

\def\tr{{\rm tr} \,}

\def\pslash{p \hspace{-1.7mm}/}

\def\uslash{u \hspace{-1.7mm}/}
\def\qslash{q \hspace{-1.7mm}/}

\def\wslash{w \hspace{-1.7mm}/}
\def\vslash{v \hspace{-1.7mm}/}

\def\w2{\tilde w^2}
\def\ws2{1}
\def\wws{(\tilde w\cdot \hat{v})}
\def\uw{(u\cdot \tilde w)}
\def\uws{(u\cdot \hat{v})}

\begin{document}
\title{Covariant and self consistent vertex corrections  \\ for pions and isobars in nuclear matter}
\author{C.L.\ Korpa,}
\affiliation{Department of Theoretical Physics, University of
Pecs, \\Ifjusag u.\ 6, 7624 Pecs, Hungary}
\author{M.F.M. Lutz and F.\ Riek\footnote{Current affiliation: Cyclotron Institute and Physics Department, \\ Texas A\&M University, College Station, Texas 77843-3366, U.S.A.} }
\affiliation{Gesellschaft f\"ur Schwerionenforschung (GSI),\\
Planck Str. 1, 64291 Darmstadt, Germany}
\date{\today}
\begin{abstract}
We evaluate the pion  and isobar propagators in cold nuclear matter self consistently
applying a covariant form of the isobar-hole model. Migdal's vertex
correction effects are considered systematically in the absence of phenomenological soft
form factors. Saturated nuclear matter is modeled by scalar and vector mean  fields for the nucleon.
It is shown that the short-range dressing of the $\pi N\Delta $ vertex has a significant
effect on the pion and isobar properties. Using realistic parameters sets we predict a
downward shift of about 50 MeV for the $\Delta$ resonance at nuclear saturation density.
The pionic soft modes are much less pronounced than in previous studies.
\end{abstract}

\maketitle

\section{Introduction}

The theoretical approaches for the nuclear pion dynamics \cite{Campbell,Oset:Weise:76,Migdal:1978, Oset:Weise,Dyugaev,Dmitriev:Suzuki,Oset:Salcedo,Migdal,Herbert:Wehrberger:Beck,Carrasco:Oset,Nieves:Oset:Recio,
Xia:Siemens:Soyeur,Arve:Helgesson,Korpa:Malfliet,Schramm,Rapp,Nakano,Lutz:Migdal,Korpa:Lutz:04,Post:Leupold:Mosel,Knoll,KoDi04,Hees:Rapp}
followed so far can be put into different categories. All works acknowledge and consider the important role of short-range correlation effects.
However, there is no common consensus about their absolute strength. The latter depends decisively on the subtle details of
the considered approach. For instance the non-relativistic computations
\cite{Nieves:Oset:Recio,Arve:Helgesson,Rapp} obtain contrasted results for the pion properties in cold nuclear matter.
With few exceptions \cite{Dmitriev:Suzuki,Herbert:Wehrberger:Beck,Schramm,Lutz:Migdal,Korpa:Lutz:04}
non-relativistic many-body techniques are applied. Also works that incorporate the feedback effect of a dressed pion propagator,
that depends sensitively on the isobar propagator itself, on the isobar self energy are in the minority
\cite{Dyugaev,Xia:Siemens:Soyeur,Korpa:Malfliet,Korpa:Lutz:04,Post:Leupold:Mosel,Knoll,Hees:Rapp}. It has been found that self-consistency is
a crucial effect for the nuclear $\pi N \Delta$ dynamics.  Moreover, the
in-medium isobar propagator should be used in the computation of the isobar-hole contribution
building up the short range correlation effects \cite{Arve:Helgesson,Korpa:Lutz:04,Post:Leupold:Mosel}.
In the early works that addressed self consistency issues \cite{Xia:Siemens:Soyeur,Korpa:Malfliet}
a quite soft phenomenological form factor was used. This implies a strong and artificial suppression
of pionic soft modes with large momentum that dominate the isobar width \cite{Korpa:Lutz:04}. The use
of such soft form factors explains why in \cite{Xia:Siemens:Soyeur,Korpa:Malfliet} quite
conventional isobar properties \cite{Hirata:Koch:Lenz:Monitz} were obtained without the
inclusion of vertex correction effects in the isobar self energy. The use of soft form factors suppresses the  in-medium mass and
width shifts of the isobar significantly. Noteworthy is the work  \cite{Arve:Helgesson}, in which
isobar properties were computed without relying on soft form factors. The important role of a hard factors in the nuclear $\pi N \Delta$ dynamics
was pointed out in \cite{Dyugaev}. One may conclude that a description of isobar properties
\cite{Xia:Siemens:Soyeur,Korpa:Malfliet,Korpa:Lutz:04,Knoll,Post:Leupold:Mosel,Hees:Rapp}
that relies on soft form factors should not be considered microscopic unless one includes a
strong density, energy and momentum dependence in the form factor.

A further possibly important aspect is the splitting of the isobar modes in nuclear matter
\cite{Oset:Salcedo,Arve:Helgesson,Korpa:Lutz:04}. An isobar moving through nuclear matter
manifests itself in terms of longitudinal and transverse modes described by distinct spectral
functions. The splitting of the two modes was found to be small in \cite{Oset:Salcedo,Korpa:Lutz:04}.
In contrast, in \cite{Arve:Helgesson} sizeable
effects were found depending, however, on the precise structure of the form factors used.
Notwithstanding, further clarification on the form of the isobar self energy in nuclear matter
is needed. This is of particular relevance for instance in applications to
heavy-ion reactions \cite{Hees:Rapp}.

Recently it was demonstrated \cite{Lutz:Migdal} that a covariant form of the
isobar-hole model differs significantly from non-relativistic versions thereof
\cite{Migdal,Oset:Weise,Dmitriev:Suzuki}. Relativistic corrections are
not important everywhere in phase space. As a striking example
recall the behavior of the nucleon-hole contribution to the pion self energy. A proper relativistic treatment leads to a result
proportional to $\omega^2- \vec q\,^2$, with the pion energy and momentum
$\omega$ and $\vec q $ respectively \cite{Dmitriev:Suzuki,Migdal,Lutz:Migdal}. In contrast,
a non-relativistic evaluation provides a factor $\vec q\,^2$ only \cite{Oset:Weise}.
Obviously, the non-relativistic expression is justified only in a small subspace of phase space.
Paying contribute to this observation various prescriptions
(see e.g. \cite{Arve:Helgesson}) were suggested in the literature. One may
speculate that the incompatible treatment of such effects is an important source for
conflicting sets of Migdal parameters used in the literature
\cite{Dmitriev:Suzuki,Arve:Helgesson,Nakano,Post:Leupold:Mosel}.

Though it should be possible to incorporate relativistic effects in a perturbative
manner with possible partial summations required, we argue that it is more economical
to perform computations that are strictly covariant.
Applying the projector techniques
developed recently \cite{Lutz:Kolomeitsev,Lutz:Korpa:02,Lutz:Migdal,Lutz:Korpa:Moeller:2007} it is straight forward to perform
such calculations. In \cite{Korpa:Lutz:04} a first manifest
covariant and self consistent computation of the pion self energy was presented. The incorporation of scalar and vector mean fields for the
nucleon was worked out recently in \cite{Lutz:Korpa:Moeller:2007} at hand of the nuclear antikaon dynamics.

The purpose of this work is to extend the previous studies of two of us \cite{Lutz:Migdal,Korpa:Lutz:04}.
We compute the isobar self energy in a covariant and self consistent manner generalizing the
covariant isobar-hole model of \cite{Lutz:Migdal}. Vertex correction effects
as well as the longitudinal and transverse isobar modes are treated consistently.
Results will be presented for a range of parameters centered around a parameter set that was
found to be compatible in \cite{Riek:Lutz:Korpa:2008} with the nuclear photoabsorption data \cite{photo-absorption}.
The effect of various approximation is discussed and illustrated comprehensively.

\section{Covariant isobar-hole model }

We specify the isobar-hole model in its covariant form \cite{Nakano,Lutz:Migdal}.
The interaction  of pions with nucleons and isobars is modelled by
the leading order vertices
\begin{eqnarray}
{\mathcal L}&=& \frac{f_{N}}{m_\pi}\,\bar \psi
\,\gamma_5\,\gamma^\mu\,(\partial_\mu \vec \pi \,)\,\vec \tau
\,\psi+ \frac{f_{\Delta }}{m_\pi} \,\Big( \bar \psi^\mu
\, (\partial_\mu \vec \pi\,)\,\vec T\,\psi + {\rm h.c.} \Big) \,,
\label{pin-pid}
\end{eqnarray}
where we use $T^\dagger_i\,T_j = \delta_{ij}- \tau_i\,\tau_j /3$ and $f_{N}=0.988$ and
$f_{\Delta }= 1.85 $ in this work.
Short range correlation effects are modelled using
the covariant forms of the Migdal interaction vertices as introduced in
\cite{Nakano,Lutz:Migdal}
\begin{eqnarray}
{\mathcal L}_{\rm Migdal} &=&g_{11}'\, \frac{f^2_{N}}{m^2_\pi} \,
\Big(\bar  \psi \,\gamma_5\,\gamma_\mu\,\vec \tau \,\psi \Big)
\,\Big(\bar  \psi \,\gamma_5 \,\gamma^\mu\,\vec \tau \,\psi\Big)
\nonumber\\
&+& g_{22}'\, \frac{f^2_{\Delta }}{m^2_\pi} \, \Bigg( \Big(\bar
\psi_\mu \,\vec T \,\psi \Big) \,\Big(\bar \psi \,\vec T^\dagger
\psi^\mu\Big) + \left(\Big(\bar \psi_\mu \,\vec T \,\psi \Big)
\,\Big(\bar \psi^\mu \,\vec T \,\psi\Big)+{\rm h.c.}\right) \Bigg)
\nonumber\\
&+& g_{12}'\, \frac{f_{N}\,f_\Delta}{m^2_\pi} \,\Big(\bar \psi
\,\gamma_5\,\gamma_\mu\,\vec \tau \,\psi \Big) \left( \Big(\bar
\psi^\mu \, \vec T \,\psi\Big) +{\rm h.c.}\right)
\,,\label{cov-Migdal}
\end{eqnarray}
where it is understood that the local vertices are to be used at
the Hartree level. The Fock contribution can be cast into the form
of a Hartree contribution by a simple Fierz transformation.
Therefore it only renormalizes the coupling strength in
(\ref{cov-Migdal}) and can be omitted here. The Lagrangian densities (\ref{pin-pid}, \ref{cov-Migdal})
are effective in the sense that we consider their coupling constants as functions of the
nuclear density. This is justified since we do not incorporate the physics of higher lying
baryon resonances nor further mesonic degrees of freedom like the vector mesons explicitly. Integrating out
more massive degrees freedom leads to a density dependence of the coupling constants necessarily, which however,
is expected to be quite smooth due to high-mass nature of the modes treated implicitly.

Unfortunately, there is yet no set
of Migdal parameters universally accepted. For instance, the computation
\cite{Arve:Helgesson} used the universal values $g_{11}'=g_{12}'=g_{22}'=0.60$,
based on a study of isobar properties. Universal values for the Migdal parameters were
suggested first in \cite{Campbell}. Nakano et al.
\cite{Nakano} deduce the constraint
$ g_{11}'= 0.585 $ together with $g_{12}'= 0.191+ 0.051 \,g_{22}' $
insisting on the empirical quenching factor $Q=0.9$ of the Gamow-Teller resonance
\cite{Wakasa}. Their consideration assumes that the quenching
results exclusively from a mixing of the nucleon-hole and the isobar-hole state.
In our work the parameters $g_{ij}'$ are varied around the values $g_{11}' \simeq 1.0, g_{22}'=g_{12}' \simeq 0.4 $
obtained from a detailed analysis of the nuclear photo absorption data \cite{Riek:Lutz:Korpa:2008}.

Our studies will be based on the in-medium nucleon propagator parameterized in terms of scalar
and vector mean fields:
\begin{eqnarray}
&& {\mathcal S}(p,u) = \frac{1}{\pslash-\Sigma_V^N\,\uslash -m_N + i\,\epsilon} + \Delta S(p,u)\,,
\qquad  m_N =m_N^{\rm vac}- \Sigma_N^S \,,
\nonumber\\
&& \Delta S (p,u) = 2\,\pi\,i\,\Theta \Big[p \cdot u-\Sigma_V^N \Big]\,
\delta\Big[(p-\Sigma^N_V\,u)^2-m_N^2\Big]\,
\nonumber\\
&& \qquad \qquad \times \,\Big( \pslash- \Sigma_V\,\uslash +m_N \Big)\,\Theta \Big[k_F^2+p^2-(u\cdot p)^2\,\big]\,,
\label{def-SN}
\end{eqnarray}
where the Fermi momentum $k_F$ specifies the nucleon density $\rho$ with
\begin{eqnarray}
\rho = -2\,\tr \,\gamma_0\,\int \frac{d^4p}{(2\pi)^4}\,i\,\Delta S(p,u)
= \frac{2\,k_F^3}{3\,\pi^2\,\sqrt{1-\vec u\,^2/c^2}}  \;.
\label{rho-u}
\end{eqnarray}
In the rest frame of the bulk with $u_\mu=(1,\vec 0\,)$ one recovers with (\ref{rho-u}) the
standard result $\rho = 2\,k_F^3/(3\,\pi^2)$. We assume isospin symmetric nuclear matter.

The focus of our work is the study of the in-medium isobar propagator
$S_{\mu \nu}(w,u)$, the solution of Dyson's equation
\begin{eqnarray}
\!\!&&S^{\mu \nu}_0(w) =\frac{-1}{\wslash-m_\Delta+ i\,\epsilon} \left(
g^{\mu \nu}-\frac{\gamma^\mu\,\gamma^\nu }{3} -\frac{2\,w^\mu
\,w^\nu }{3\,m_\Delta^2} -\frac{\gamma^\mu \,w^\nu -w^\mu\,
\gamma^\nu}{3\,m_\Delta} \right)\,,
\nonumber\\
&& S_{\mu \nu} (w,u) = S^{(0)}_{\mu \nu}(w-\Sigma^\Delta_V\,u)
+ S^{(0)}_{\mu \alpha}(w-\Sigma^\Delta_V\,u)\,\Sigma^{\alpha \beta}(w,u)\, S_{\beta \nu}(w,u)\,,
\label{Dyson}
\end{eqnarray}
where we allow for a vector mean field of the isobar.

\begin{figure}[t]
\begin{center}
\vskip-0.1cm
\hskip-0.3cm
\parbox{8.2cm}{\includegraphics[clip=true,width=8.2cm]{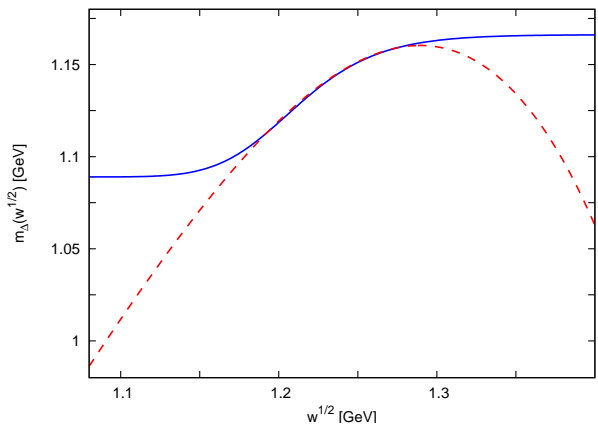}}
\parbox{5cm}{\vspace*{-0.5cm}
\caption{Phenomenological isobar mass functions $m^{\rm vac}_\Delta (\sqrt{w^2}\,)$ that lead to
the reproduction of the P33 pion-nucleon scattering amplitude. The dashed (solid)
line leads to an exact (approximate) reproduction of the amplitude.}
\label{fig:1} }
\end{center}
\end{figure}

In nuclear matter the isobar self energy tensor, $\Sigma_{\mu \nu}(w,u)$, is
a quite complicated object which involves the time-like 4-vector $u_\mu$
characterizing the nuclear matter frame. In order to arrive at a reproduction
of the P33 pion-nucleon partial-wave amplitude we allow for a phenomenological energy
dependence in the free-space isobar mass. We write
\begin{eqnarray}
m_\Delta=m^{\rm vac}_\Delta(\sqrt{w^2}\,)-\Sigma_S^\Delta \,,
\label{def-sigma-s-delta}
\end{eqnarray}
where we introduce also a scalar mean field $\Sigma_S^\Delta$ for the isobar. At nuclear saturation density
we found the  values $\Sigma^\Delta_V \simeq -0.25$ GeV and $\Sigma^\Delta_S\simeq -0.11$ GeV to be consistent
with the nuclear photo absorption data in an application of the present covariant and self consistent
many-body approach \cite{Riek:Lutz:Korpa:2008}. Note that latter values are scheme dependent reflecting the
particular in-medium processes taken into account explicitly.

Making the assumption that the P33 amplitude is determined completely by
the s-channel exchange of the dressed isobar, for a given isobar self energy the mass function
$m^{\rm vac}_\Delta$ can be expressed directly in terms of the empirical P33 phase shift. Based on the self
energy to be specified in section 5 we arrive at the mass function shown in Fig. 1 with a dashed line.
The sizeable variation of $m^{\rm vac}_\Delta$ on $\sqrt{w^2}$ reflects contributions to the P33 amplitude that
are characterized by left-hand branch points. The amplitude receives, besides the s-channel isobar exchange, additional
contributions like from the nucleon u-channel process. Since the latter contribution will be considered being,
implied by (\ref{pin-pid}), it is not consistent to proceed with the dashed mass function of Fig. 1.
A fully consistent approach would require at least the unitarization of the sum of s-channel isobar
and u-channel nucleon exchange processes. This is, however, beyond the scope of the present work.
In order to correct for the presence of the u-channel nucleon exchange we determine the phenomenological mass
function $m^{\rm vac}_\Delta(\sqrt{w^2})$ in the following way: the s-channel isobar contribution is
adjusted to reproduce the imaginary part of the P33 partial wave amplitude  in the vicinity of the
isobar peak. Away from the resonance the mass function is kept constant. The result is shown in Fig. 1 with
a solid line.  As compared to the dashed line, which reproduces the P33 amplitude exactly,
the solid line shows a much reduced variation. This is welcome since the smoother the phenomenological mass function
the smaller are the uncertainties implied by the ansatz (\ref{def-sigma-s-delta}).

The quality of our prescription is illustrated in Fig. 2, where the empirical
P33 partial wave amplitude in the convention of \cite{Korpa:Lutz:04}
is confronted with the phenomenological amplitude. Real and imaginary parts agree well in  the
resonance region. Significant deviations are noted close to threshold only, where we expect a strong energy
dependence from the u-channel contributions. This is confirmed by the additional solid line of Fig. 2 which  shows
the contribution of the u-channel nucleon exchange process. Close to threshold it is largest almost
making up the difference of the empirical and phenomenological amplitude.

\begin{figure}[t]
\begin{center}
\vskip-0.1cm
\hskip-0.3cm
\parbox{8.2cm}{\includegraphics[clip=true,width=8.2cm]{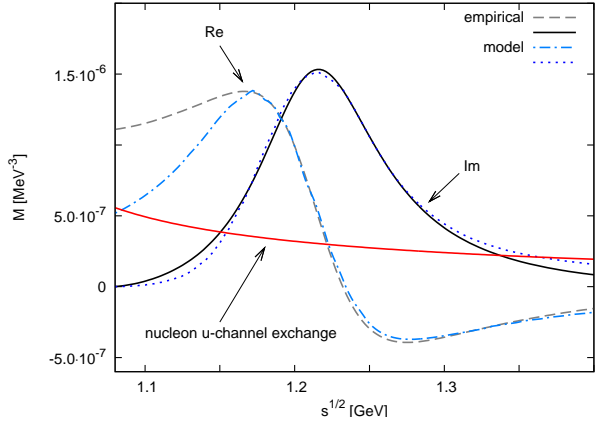}}
\parbox{5cm}{\vspace*{-0.5cm}
\caption{Comparison of the empirical P33 amplitude from \cite{Arndt} with the phenomenological
one  as implied by (\ref{def-sigma-s-delta}) with the solid mass function of Fig. 2. The additional solid
line is the contribution of the u-channel nucleon exchange process.}
\label{fig:2} }
\end{center}
\end{figure}

\section{Pion self energy}

In this section we evaluate the pion self energy as
implied by the interaction (\ref{pin-pid}) for a given isobar propagator $S_{\mu \nu}(w,u)$.
The latter will be specified in subsequent sections. The central  objects
to compute are the nucleon- and isobar-hole loop tensors, $\Pi^{(N h)}_{\mu \nu}(q,u)$
and $\Pi^{(\Delta h)}_{\mu \nu}(q,u)$, which we define by
\begin{eqnarray}
&& \! \Pi_{\mu \nu}^{(\Delta h)}(q,u) = \frac{4}{3}\,
\frac{f^2_{\Delta }}{m^2_\pi} \int \frac{d^4 p}{(2\pi)^4}\,i\,\tr
\,\Delta S(p,u)\,  S_{\mu \nu }(p+q,u) + (q_\mu
\to -q_\mu )  \,, \nonumber\\
&& \!  \Pi_{\mu \nu}^{(Nh)}(q,u) = 2\,\frac{f^2_{N}}{m^2_\pi}
\,\int \frac{d^4 p}{(2\pi)^4}\,i\,\tr \,\Bigg( \Delta S(p,u)\,
\gamma_5 \,\gamma_\mu \,\frac{1}{\pslash - \Sigma_V^N\,\uslash+\qslash -m_N}
 \,\gamma_5 \,\gamma_\nu \nonumber\\
&& \qquad \qquad \qquad \; + \frac{1}{2}\, \Delta S(p,u)\,
\gamma_5 \,\gamma_\mu \,\Delta S(p+q,u) \,\gamma_5 \,\gamma_\nu
\Bigg)
 + (q_\mu
\to -q_\mu ) \,,  \label{nh-dh}
\end{eqnarray}
with the isobar propagator, $S_{\mu \nu}(w,u)$ of (\ref{Dyson}), and
the in-medium  part of the nucleon propagator, $\Delta S (p,u)$ as specified in (\ref{def-SN}).

The computation of short range correlation effects is considerably streamlined
upon decomposing the nucleon- and isobar hole tensors,
\begin{eqnarray}
&& \Pi_{\mu \nu}^{(N h)}(q,u) = \sum_{i,j=1}^2\,
\Pi_{ij}^{(N h)}(q,u)\,L_{\mu \nu}^{(ij)}(q,u)+ \Pi_{\,T}^{(N h)}(q,u)\,T_{\mu \nu}(q,u) \,,
\nonumber\\
&& \Pi_{\mu \nu}^{(\Delta h)}(q,u) = \sum_{i,j=1}^2\,
\Pi_{ij}^{(\Delta h)}(q,u)\,L_{\mu \nu}^{(ij)}(q,u)
+ \Pi_{\,T}^{(\Delta h)}(q,u)\,T_{\mu \nu}(q,u)\,,
\label{def-decom-pi}
\end{eqnarray}
in terms of a complete set of Lorentz structures $L_{\mu \nu}^{(ij)}(q,u)$ and
$T_{\mu \nu}(q,u)$. A convenient basis that
enjoys projector properties was suggested in \cite{Lutz:Migdal}. We recall the
definitions
\begin{eqnarray}
&& L^{(22)}_{\mu \nu}(q,u)=
\Big[ \frac{(q \cdot u)}{q^2}\,q_\mu-u_\mu \Big]\,
\frac{q^2}{q^2-(q \cdot u)^2}\,
\Big[ \frac{(q \cdot u)}{q^2}\,q_\nu-u_\nu \Big] \,,
\nonumber\\
&& L^{(12)}_{\mu \nu}(q,u)=L^{(21)}_{\nu \mu}(q,u)=q_\mu\,\sqrt{\frac{1}{q^2-(q \cdot u)^2}}\,
\Big[ \frac{(q \cdot u)}{q^2}\,q_\nu-u_\nu \Big]\,,
\nonumber\\
&& L_{\mu \nu}^{(11)}(q,u) = \frac{q_\mu\,q_\nu}{q^2} \,,
\qquad T_{\mu \nu}(q,u) = g_{\mu \nu}- \frac{q_\mu\,q_\nu}{q^2}- L^{(22)}_{\mu \nu}(q,u) \,.
\label{add22b}
\end{eqnarray}
The presentation of explicit expressions for the longitudinal and transverse
nucleon- and isobar-hole loop functions
is relegated to the Appendix A. The latter follow by simple contraction of
the tensors $\Pi_{\mu \nu}(q,u)$ with the projectors in (\ref{add22b}). The results
depend on the details of the in-medium isobar propagator which will be specified in
section 5 and 6.

Following \cite{Lutz:Migdal} we construct the pion self energy in terms of the
longitudinal nucleon- and isobar-hole loop functions. The self energy
can be cast into the form of a sum of $11$, $33$ and
$13$, $31$ components of an appropriate 4$\times$4 matrix,
\begin{eqnarray}
&& \Pi(q,u) = - 4\,\pi\,(1+ \frac{m_\pi}{m_N})\,b_{\rm eff}\,\rho
\nonumber\\
&& \quad -q^2\,\Big[\Big( 1-\Pi^{(L)}\,g^{(L)}\Big)^{-1} \,\Pi^{(L)} \Big]_{11} -
q^2\,\Big[\Big( 1-\Pi^{(L)} \,g^{(L)}\Big)^{-1} \,\Pi^{(L)}  \Big]_{33}
\nonumber\\
&& \quad -q^2\,\Big[\Big( 1-\Pi^{(L)} \,g^{(L)}\Big)^{-1} \,\Pi^{(L)} \Big]_{13}
- q^2\,\Big[\Big( 1-\Pi^{(L)}\,g^{(L)} \Big)^{-1}\,\Pi^{(L)}  \Big]_{31} \,,
\label{full-result}
\end{eqnarray}
where
\begin{eqnarray}
&& g^{(L)} = \left(
\begin{array}{llll}
g_{11}' & 0 & g_{12}' & 0 \\
0 & g_{11}' & 0 & g_{12}'  \\
g_{12}' & 0 & g_{22}' & 0 \\
0 & g_{12}' & 0 & g_{22}'
\end{array}
\right) \,, \quad \!\Pi^{(L)}= \left(
\begin{array}{llll}
\Pi_{11}^{(N h)} & \Pi_{12}^{(N h)} & 0 & 0 \\
\Pi_{21}^{(N h)} & \Pi_{22}^{(N h)} & 0 & 0  \\
0 & 0 & \Pi_{11}^{(\Delta h)} & \Pi_{12}^{(\Delta h)} \\
0 & 0 & \Pi_{21}^{(\Delta h)} & \Pi_{22}^{(\Delta h)}
\end{array}
\right)\,. \label{def-matrix}
\end{eqnarray}

In (\ref{full-result}) we allow for a background term linear in the nuclear density reflecting a s-wave
pion-nucleon interaction. Such a term is motivated by the fact that the vertices of (\ref{pin-pid}) do not reproduce the
empirical s-wave scattering pion-nucleon length. At tree-level the vertices (\ref{pin-pid})
lead to a pion-nucleon isospin averaged scattering length of the form \cite{Lutz:Kolomeitsev},
\begin{eqnarray}
4\,\pi\,(1+ \frac{m_\pi}{m_N})\,a_{\pi N} =- \frac{f_N^2}{m_N} - \frac{8}{9}\,
\frac{f_\Delta^2}{m_\Delta}\,
\Big( 1+2\,\frac{m_N}{2\,m_\Delta}\Big) \,.
\label{apin}
\end{eqnarray}
This leads to $a_{\pi N} \simeq -0.09$ fm, a significant overestimation of the empirical scattering length of about
$  -0.01 $ fm \cite{Lutz:Kolomeitsev}. Using the unitarized isobar propagator as implied by the one-loop isobar self energy of
section 5 we obtain $a_{\pi N} \simeq + 0.00$ fm instead, a value significantly reduced and closer to the empirical constraint.
In order to correct for the remaining slight mismatch we use $b_{\rm eff}\simeq  -0.01$ fm in (\ref{full-result}).

There are two important technical issues we need to emphasize here. First the application of the longitudinal and
transverse projectors in (\ref{def-decom-pi}) implies that the loop functions have to satisfy specific
constraint conditions. They follow from the observation that the polarization  tensor $\Pi_{\mu \nu}(q,u)$ is regular, in
particular at $q^2 =0$ and at $q^2= (q \cdot u)^2$. It must hold
\begin{eqnarray}
&&\Pi_{22}(q,u)= \Pi_{11}(q,u)-i\,\Pi_{12}(q,u)-i\,\Pi_{21}(q,u)+ {\mathcal O} \left( q^2\right) \,,
\nonumber\\
&& \Pi_{22}(q,u)= \Pi_T(q,u)+ {\mathcal O} \left( (q \cdot u)^2-q^2\right) \,.
\label{constraint-polarization}
\end{eqnarray}
The reader may wonder why we discuss this point. After all the integrals (\ref{nh-dh}) are finite and the conditions
(\ref{constraint-polarization}) should be satisfied automatically. However, we argue in favor of a finite renormalization which
is not necessarily compatible with (\ref{constraint-polarization}). A finite renormalization of the isobar-hole loop functions
is useful as to suppress the formation of ghosts in the pion self energy. The latter may be absorbed into a redefinition
of the Migdal's short-range interaction (\ref{cov-Migdal}). The occurrence of ghost causes a
severe problem, in particular in a self consistent approach. It implied that the pion self energy does
not satisfy a Lehman representation anymore. A ghost state is present if the pion self energy has a pole for complex energies, i.e.
\begin{eqnarray}
D(\omega )=\det [1- \Pi^{(L)}(\omega,\vec q\,)\,g^{L}] = 0   \qquad {\rm with} \qquad \Im \omega  \neq 0\,.
\end{eqnarray}
Note that a function that satisfies a Lehman representation can have poles only on the 2nd or higher Riemann sheets.
In fact, we observe that such artifacts are avoided typically once a finite renormalization is implemented such that
all elements $\Pi_{ij}(\omega, \vec q\,)$ are bounded for large energies, i.e.
\begin{eqnarray}
\lim_{\omega \to \pm \infty }|\Pi_{ij}(\omega, \vec q\,)| < \infty \,.
\label{bounded-assumption}
\end{eqnarray}
As detailed in Appendix A we introduce a finite renormalization for the isobar-hole loop functions by
insisting on subtracted dispersion-integral representations thereof.  The construction of the latter was determined by the constraints (\ref{constraint-polarization}). We checked that our numerical pion self energies  satisfy  a once-subtracted
dispersion-integral representation to reasonable accuracy.
In our self consistent simulations we impose such a dispersion-integral representation, where the
subtraction constant is determined as to find agreement with the direct computation at $\omega^2 -\vec q\,^2 =m_\pi$.

It is evident that there is a self-consistency issue here. The isobar
propagator defining the isobar-hole loop functions in (\ref{nh-dh}) is a crucial
ingredient to evaluate the pion self energy. Since, the
isobar self energy depends sensitively on the pion propagator itself a self consistent
computation is required. The importance of self consistency, as discussed
above, will be addressed in the second last section when presenting numerical results.

\section{Isobar propagator}

The solution of the
Dyson equation (\ref{Dyson}) requires a detailed study of
the Lorentz-Dirac structure of the isobar propagator.
Consider the propagator, $ S_{\mu \nu }(w,u) $, in the nuclear medium. From covariance
we expect a general decomposition of the form,
\begin{eqnarray}
S^{\mu \nu}(w, u) = \sum_{i,j}\,S^{(p)}_{[ij]}(v,u)\,P_{[ij]}^{\mu
\nu}(v,u) +\sum_{i,j}\,S^{(q)}_{[ij]}(v,u)\,Q_{[ij]}^{\mu
\nu}(v,u) \,, \label{S-decompose}
\end{eqnarray}
in terms of invariant functions, $S_{[ij]}^{(p,q)}(v,u)$, and a
complete set of Dirac-Lorentz tensors $P^{\mu \nu}_{[ij]}(v,u)$
and $Q^{\mu \nu}_{[ij]}(v, u)$. For latter convenience we
introduce
\begin{eqnarray}
v_\mu= w_\mu - \Sigma_V^N\,u_\mu \,. \label{def-vmu}
\end{eqnarray}
A suitable basis was constructed in
\cite{Lutz:Kolomeitsev,Lutz:Korpa:02}, enjoying the projector
properties
\begin{eqnarray}
&& Q_{[ik]}^{\mu \alpha }\,g_{\alpha \beta}\,P_{[lj]}^{\beta \nu }
= 0 = P_{[ik]}^{\mu \alpha }\,g_{\alpha \beta}\,Q_{[lj]}^{\beta \nu }
\;, \quad
\nonumber\\
&& Q_{[ik]}^{\mu \alpha }\,g_{\alpha \beta}\,Q_{[lj]}^{\beta \nu }
= \delta_{kl}\,Q_{[ij]}^{\mu \nu} \;,\quad
P_{[ik]}^{\mu \alpha }\,g_{\alpha \beta}\,P_{[lj]}^{\beta \nu }
= \delta_{kl}\,P_{[ij]}^{\mu \nu}\;.
\label{proj-algebra}
\end{eqnarray}
This particular basis streamlines the computation of the in-medium part of
the isobar self energy significantly. It was applied also in \cite{KoDi04}.
In particular the algebra (\ref{proj-algebra})
illustrates the decoupling of helicity one-half (p-space) and
three-half modes (q-space). Decomposing the isobar self energy
\begin{eqnarray}
\Sigma^{\mu \nu}(w,u)
=\sum_{i,j}\,\Sigma^{(p)}_{[ij]}(v,u)\,P_{[ij]}^{\mu \nu}(v,u)
+\sum_{i,j}\,\Sigma^{(q)}_{[ij]}(v,u)\,Q_{[ij]}^{\mu \nu}(v,u) \,,
\label{Sigma-decompose}
\end{eqnarray}
into the set of projectors it is straightforward to evaluate the isobar propagator. The Dyson
equation (\ref{Dyson}) maps onto two simple matrix equations. First, the bare propagator
\begin{eqnarray}
S_0^{\mu \nu }(w)=\sum_{i,j}\,S^{(p)}_{0,[ij]}(v,u)\,P_{[ij]}^{\mu
\nu}(v,u) +\sum_{i,j}\,S^{(q)}_{0,[ij]}(v,u)\,Q_{[ij]}^{\mu
\nu}(v,u) \,, \label{S0-decompose}
\end{eqnarray}
needs to be decomposed in terms of the projectors. Second, the
six-dimensional matrix $\Sigma^{(p)}(v,u)$ and two-dimensional
matrix $\Sigma^{(q)}(v,u)$ have to be evaluated. The final form of
the isobar propagator, specified in terms of the invariant
matrices $S^{(p)}(v,u)$ and $S^{(q)}(v,u)$, follows by simple matrix
manipulations
\begin{eqnarray}
&& S^{(p)}(v,u)=S_0^{(p)}(v,u)\,\Big[1-
\Sigma^{(p)}(v,u)\,S_0^{(p)}(v,u) \Big]^{-1}\,,
\nonumber\\
&& S^{(q)}(v,u)=S_0^{(q)}(v,u)\,\Big[1-
\Sigma^{(q)}(v,u)\,S_0^{(q)}(v,u) \Big]^{-1} \,.
\label{isobar-propagator-final}
\end{eqnarray}
The transparent expressions
(\ref{isobar-propagator-final}) rely on the explicit availability
of the projector algebra. In order to keep this work self
contained we review briefly the set of projectors $P^{\mu
\nu}_{[ij]}(v,u)$ and $Q^{\mu \nu}_{[ij]}(v,u)$ introduced in
\cite{Lutz:Korpa:02}. It is convenient to express the latter in
terms of appropriate building blocks  $P_\pm$, $U_\pm$, $V_\mu$
and $L_\mu, R_\mu$ of the form:
\begin{eqnarray}
&& P_\pm(v) = \frac{1}{2}\left( 1\pm
\frac{\vslash}{\sqrt{v^2}}\right)\, ,\quad U_\pm (v,u)=P_\pm(v)
\,\frac{-i\,\gamma \cdot u}{\sqrt{(v\cdot
u)^2/v^2-1}}\,P_\mp(v)\;,
\nonumber\\
&&V_\mu (v)=\frac{1}{\sqrt{3}}\,\Big( \gamma_\mu
-\frac{\vslash}{v^2}\,v_\mu \Big) \;,\quad X_\mu(v,u)=
\frac{(v\cdot u)\,v_\mu-v^2\,u_\mu}{v^2\,\sqrt{(v \cdot
u)^2/v^2-1}} \;,
\nonumber\\
&&R_\mu (v,u) = +\frac{1}{\sqrt{2}}\,\Big(
U_+(v,u)+U_-(v,u)\Big)\,V_\mu(v)-i\,\sqrt{\frac{3}{2}}\,X_\mu(v,u)
\, , \quad
\nonumber\\
&& L_\mu(v,u) =+\frac{1}{\sqrt{2}}\,V_\mu(v)\, \Big(
U_+(v,u)+U_-(v,u)\Big) -i\,\sqrt{\frac{3}{2}}\,X_\mu(v,u) \;.
\label{def-basic}
\end{eqnarray}
For a compilation of useful properties of the building blocks $P_\pm$, $U_\pm$,
$V_\mu$ and $R_\mu, L_\mu$ we refer to the original work \cite{Lutz:Korpa:02}.
The q-space projectors are
\begin{eqnarray}
&& Q_{[11]}^{\mu \nu } =\Big( g^{\mu \nu}-\hat v^\mu\,\hat v^\nu
\Big) \,P_+ - V^\mu\,P_-\,V^\nu -L^\mu\,P_+\,R^\nu \;,
\nonumber\\
&& Q_{[22]}^{\mu \nu } =\Big( g^{\mu \nu}-\hat v^\mu\,\hat v^\nu
\Big) \,P_- - V^\mu\,P_+\,V^\nu -L^\mu\,P_-\,R^\nu \;,
\nonumber\\
&& Q_{[12]}^{\mu \nu }  = \Big(g^{\mu \nu}-\hat v^\mu\,\hat v^\nu
\Big)\,U_+ +{\textstyle{1\over 3}}\,V^\mu\,U_-\,V^\nu
\nonumber\\
&&\qquad +{\textstyle{\sqrt{8}\over 3}}\,
\Big( L^\mu\,P_+\,V^\nu +V^\mu\,P_-\,R^\nu \Big) -{\textstyle{1\over 3}}\,L^\mu\,U_+\,R^\nu\;,
\nonumber\\
&&Q_{[21]}^{\mu \nu } = \Big(g^{\mu \nu}-\hat v^\mu\,\hat
v^\nu\Big)\,U_- +{\textstyle{1\over 3}}\,V^\mu\,U_+\,V^\nu
\nonumber\\
&&\qquad +{\textstyle{\sqrt{8}\over 3}}\,
\Big( L^\mu\,P_-\,V^\nu +V^\mu\,P_+\,R^\nu \Big)-{\textstyle{1\over 3}}\,L^\mu\,U_-\,R^\nu\;,
\label{q-space-def}
\end{eqnarray}
where $\hat v_\mu = v_\mu /\sqrt{v^2} $. It is straightforward to
verify (\ref{proj-algebra}). Using the properties of the building
blocks $P_\pm$, $U_\pm$, $V_\mu$ and $L_\mu, R_\mu$
\cite{Lutz:Korpa:02} reveals that the objects $Q^{\mu \nu}_{[ij]}$
indeed form a projector algebra.

The p-space projectors have similar transparent representations. Following \cite{Lutz:Korpa:02}
it is convenient to extend the p-space algebra including objects with one or no Lorentz
index,
\begin{eqnarray}
&&\begin{array}{llll}
P_{[11]} = P_+  \,, & P_{[12]}= U_+ \,, & P_{[21]}=U_-\,, & P_{[22]}=P_- \,, \\
P^\mu_{[31]} = V^\mu \,P_+ \,,  & P^\mu_{[32]} = V^\mu \,U_+ \;, &
\bar P^\mu_{[13]} = P_+\,V^\mu \;,  &  \bar P^\mu_{[23]} = U_-\,V^\mu \;,  \\
P^\mu_{[41]} = V^\mu \,U_- \;,  & P^\mu_{[42]} = V^\mu \,P_- \;, &
\bar P^\mu_{[14]} = U_+\,V^\mu\;,  & \bar P^\mu_{[24]} = P_-\,V^\mu\;,  \\
P^\mu_{[51]} = \hat v^\mu \,P_+ \;,  & P^\mu_{[52]} = \hat v^\mu
\,U_+ \;, &
\bar P^\mu_{[15]} = P_+\,\hat v^\mu \;,  & \bar P^\mu_{[25]} = U_-\,\hat v^\mu \;, \\
P^\mu_{[61]} = \hat v^\mu \,U_- \;,  & P^\mu_{[62]} = \hat v^\mu
\,P_- \;, &
\bar P^\mu_{[16]} = U_+\,\hat v^\mu\;, & \bar P^\mu_{[26]} = P_-\,\hat v^\mu\;, \\
P^\mu_{[71]} =   L^\mu \,P_+ \;,  & P^\mu_{[72]} =   L^\mu \,U_+ \;, &
\bar P^\mu_{[17]} = P_+\,  R^\mu \;,  & \bar P^\mu_{[27]} = U_-\,  R^\mu \;, \\
P^\mu_{[81]} =   L^\mu \,U_- \;,  & P^\mu_{[82]} =   L^\mu \,P_- \;, &
\bar P^\mu_{[18]} = U_+\,  R^\mu\;, & \bar P^\mu_{[28]} = P_-\,  R^\mu\;,
\end{array}
\nonumber\\ \nonumber\\
&& P_{[i\,j]}^{\mu \nu} = P^\mu_{[i1]}\;\bar P^\nu_{[1j]} = P^\mu_{[i2]}\;\bar P^\nu_{[2j]}\,.
\label{p-space-def}
\end{eqnarray}
In the notation of (\ref{p-space-def}) the indices
$i,j$ in (\ref{S-decompose}, \ref{Sigma-decompose}, \ref{S0-decompose}) run from
3 to 8 in the p-space. The set of identities (\ref{proj-algebra})
extends naturally
\begin{eqnarray}
&&P_{[ik]}\cdot P_{[lj]} =\delta_{kl}\,P_{[ij]} \;, \quad
P^\mu_{[ik]}\;\bar P^\nu_{[lj]}= \delta_{kl}\,P_{[ij]}^{\mu \nu}\,,\quad
\bar P^\mu_{[ik]}\,g_{\mu \nu}\,P^\nu_{[lj]}= \delta_{kl}\,P_{[ij]} \;,
\nonumber\\
&& Q_{[ik]}^{\mu \alpha }\,g_{\alpha \beta}\,P_{[lj]}^{\beta }
= 0 = \bar P_{[ik]}^{\alpha }\,g_{\alpha \beta}\,Q_{[lj]}^{\beta \nu }\;.
\label{proj-algebra-extension}
\end{eqnarray}

The algebra (\ref{p-space-def}) proves convenient in
solving various problems.
Using the projector formalism we compute the in-medium
isobar self energy as implied by the interaction vertex (\ref{pin-pid}) at
the one-loop level  in a manifest covariant fashion.

\section{Isobar self energy and pion-nucleon scattering }

It proves convenient to extract the isobar propagator from an appropriately constructed model of the
pion-nucleon scattering amplitude. Set up in this way all results are induced by expressions
already presented in \cite{Lutz:Korpa:02} upon the application of simple substitution rules.
Recall the in-medium Bethe-Salpeter equation,
\begin{eqnarray}
\!\!\!\!\!&&{\mathcal T} (\bar k ,k ;w ,u) = {\mathcal V} (\bar k ,k ;w ,u)
+\int \frac{d^4l}{(2\pi)^4}\,{\mathcal V}(\bar k , l;w,u )\,{\mathcal G}(l;w,u)\,
{\mathcal T}(l,k;w ,u )\;,
\nonumber\\
\!\!\!\!\!&& {\mathcal G}({\textstyle {1\over 2}}\,w-l;w,u)=-i\,S (w-l,u)\,
\Big[ l^2-m_\pi^2- \Pi(l,u)\Big]^{-1} \,,
 \label{BS-eq}
\end{eqnarray}
where $q,p, \bar q , \bar p$ are the initial and final pion and nucleon 4-momenta and
\begin{eqnarray}
w=p+q=\bar p+\bar q \,, \qquad k = {\textstyle{1\over 2}}\,(p-q)
\,, \qquad \bar k = {\textstyle{1\over 2}}\,(\bar p-\bar q)\,.
\end{eqnarray}
The two-particle propagator, $ {\mathcal G} (l;w,u)$, is specified in
terms of the nucleon propagator $S(p,u)$ of (\ref{def-SN})
and the pion propagator written in terms of the in-medium
self energy $\Pi(l,u)$ of (\ref{full-result}).

In order to generate the isobar self energy $\Sigma^{\mu \nu}(w,u)$,
we introduce the interaction kernel
\begin{eqnarray}
{\mathcal V}(\bar k,k;w,u)=
-\frac{f^2_\Delta}{m_\pi^2}\,\bar q_\mu\,S_0^{\mu \nu}(w- \Sigma^\Delta_V\,u)\,q_\nu \,,
\label{identify}
\end{eqnarray}
where we allow for the presence of a vector mean field.
The isospin projector is suppressed in (\ref{identify}) (see e.g. \cite{Korpa:Lutz:04}).
The particular choice (\ref{identify}) implies a scattering amplitude, which
determines the isobar propagator, $S_{\mu \nu}(w,u)$, by
\begin{eqnarray}
&& {\mathcal T}(\bar k ,k ;w ,u) =-\frac{f^2_\Delta}{m_\pi^2}\,\bar q_\mu\,S^{\mu \nu}(w,u)\,q_\nu \,.
\label{ST-match}
\end{eqnarray}
The system is solved conveniently by decomposing the interaction kernel
into a set of projectors, where we apply the
projectors constructed in terms of the 4-momentum $v_\mu= w_\mu
-\Sigma_V\,u_\mu$ and $u_\mu $ rather than $w_\mu$ and $u_\mu$:
\begin{eqnarray}
&& {\mathcal V} = \sum_{i,j} V^{(p)}_{[ij]}(v,u)\,\bar
q_\mu\,P^{\mu \nu}_{[ij]}(v,u)\,q_\nu +\sum_{i,j}
V^{(q)}_{[ij]}(v,u)\,\bar q_\mu\,Q^{\mu \nu}_{[ij]}(v,u)\,q_\nu\,.
\label{def-V-exp}
\end{eqnarray}
For the general case with $\Sigma^\Delta_V \neq \Sigma^N_V$ the
derivation of $V^{(p,q)}_{[ij]}(v,u)$ as implied by (\ref{identify} ) is somewhat
tedious though straight forward. The expressions are listed in Appendix B.
In the limit $\Sigma^\Delta_V \to \Sigma_V^N$ the expressions simplify with:
\begin{eqnarray}
&&V^{(q)}_{[11]}=V^{(p)}_{[77]}=
+\frac{f_\Delta^2}{m_\pi^2}\,\frac{1}{\sqrt{v^2}-m_\Delta}\,,
\qquad
 V^{(q)}_{[22]}=V^{(p)}_{[88]}=
-\frac{f_\Delta^2}{m_\pi^2}\,\frac{1}{\sqrt{v^2}+m_\Delta}\,,
\nonumber\\
&& V^{(p)}_{[55]}= -\frac{2}{3}\,\frac{f_\Delta^2}{m_\pi^2}\,
\frac{\sqrt{v^2}+m_\Delta}{m_\Delta^2}
 \,, \qquad   V^{(p)}_{[66]}= +\frac{2}{3}\,\frac{f_\Delta^2}{m_\pi^2}\,
\frac{\sqrt{v^2}-m_\Delta}{m_\Delta^2}
\,,
\nonumber\\
&& V^{(p)}_{[53]}=V^{(p)}_{[35]} = +
\frac{1}{\sqrt{3}}\,\frac{f_\Delta^2}{m_\pi^2}
\,\frac{1}{\,m_\Delta} \,, \qquad
 V^{(p)}_{[64]}=V^{(p)}_{[46]} =
-\frac{1}{\sqrt{3}}\,\frac{f_\Delta^2}{m_\pi^2}
\,\frac{1}{\,m_\Delta} \,,
\label{V-specify}
\end{eqnarray}
where only components that are non-zero are specified in
(\ref{V-specify}). A corresponding decomposition is implied for
the in-medium scattering amplitude
\begin{eqnarray}
&& {\mathcal T} = \sum_{i,j} T^{(p)}_{[ij]}(v,u)\,\bar
q_\mu\,P^{\mu \nu}_{[ij]}(v,u)\,q_\nu +\sum_{i,j}
T^{(q)}_{[ij]}(v,u)\,\bar q_\mu\,Q^{\mu \nu}_{[ij]}(v,u)\,q_\nu\,,
\nonumber\\
&& T^{(p)}(v,u)= V^{(p)}(v,u)\,\Big[ 1- J^{(p)}(v,u)\,V^{(p)}(v,u)
\Big]^{-1}\,,
\nonumber\\
&& T^{(q)}(v,u)= V^{(q)}(v,u)\,\Big[ 1- J^{(q)}(v,u)\,V^{(q)}(v,u)
\Big]^{-1}\,. \label{T-result}
\end{eqnarray}
The scattering amplitude ${\mathcal T}$ is determined by the
interaction kernel (\ref{V-specify}) and two matrices of loop
functions $J^{(p)}_{[ij]}(v,u)$ and $J^{(q)}_{[ij]}(v,u)$.
Comparing (\ref{T-result}) with (\ref{identify}) and
(\ref{isobar-propagator-final}) we identify
\begin{eqnarray}
&& S^{(p)}_{0,[ij]}(v,u) =
-\frac{m_\pi^2}{f_\Delta^2}\,V^{(p)}_{[ij]}(v,u) \,,\qquad
S^{(q)}_{0,[ij]}(v,u) =
-\frac{m_\pi^2}{f_\Delta^2}\,V^{(q)}_{[ij]}(v,u)
\nonumber\\
&& \Sigma^{(p)}_{[ij]}(v,u) =
-\frac{f_\Delta^2}{m_\pi^2}\,J^{(p)}_{[ij]}(v,u) \,,\qquad \;\;\,
\Sigma^{(q)}_{[ij]}(v,u) =
-\frac{f_\Delta^2}{m_\pi^2}\,J^{(q)}_{[ij]}(v,u) \,.
\label{def-sigma-identify}
\end{eqnarray}

The form of the loop functions can be taken over from \cite{Lutz:Korpa:02,Lutz:Korpa:Moeller:2007}.
The evaluation of the real parts of the loop functions requires
great care. The imaginary parts of the loop functions are unbounded at large energies. Thus
power divergencies arise if the real parts are evaluated by means
of an unsubtracted dispersion-integral ansatz. The task is to
device a subtraction scheme that avoids kinematical singularities and that
eliminates all power divergent terms systematically. The latter are unphysical and in a
consistent effective field theory approach must be absorbed into
counter terms. Only the residual strength of the counter terms may
be estimated by a naturalness assumption reliably. Since we want
to neglect such counter terms it is crucial to set up the
renormalization in a proper manner. The scheme developed in
\cite{Lutz:Korpa:Moeller:2007} avoids the occurrence of power-divergent structures and is free of
kinematical singularities.

The loop functions $J_{[ij]}^{(p,q)}(v_0\,, \vec w\,)$ are
expressed in terms of a basis spanned by 13 master loop functions,
$J_{n}(v_0\,, \vec w\,)$ as detailed in \cite{Lutz:Korpa:Moeller:2007}. We assume nuclear matter at rest
for simplicity. The
master loop functions are evaluated by a dispersion integral  of the form
\begin{eqnarray}
J_{n}(v_0,\vec w\,) = \int_{-\infty}^{+\infty} \frac{d \bar
v_0}{\pi}\, \frac{\Delta J_{n}(\bar v_0; v_0,\vec
w\,)}{\bar v_0-v_0-i\,\epsilon\,(\bar
v_0 -\mu)}
+ J^C_{n}(v_0, \vec w\,)
\,,\label{disp-integral}
\end{eqnarray}
where $\mu^2=m_N^2+k_F^2$. We introduce spectral weight functions, $\Delta J_{n}(\bar v_0; v_0,\vec w\,)$, that depend on 'external' and
'internal' energies $v_0=w_0-\Sigma_V$ and $\bar v_0$. We identify
\begin{eqnarray}
&&\Delta J_{n}(\bar v_0;v_0,\vec w\,) = \int \frac{d\,^3
l}{2\,(2\, \pi)^3}\, \,\Big(m_N^2+\vec l\,^2\,
\Big)^{-\frac{1}{2}}\,
\nonumber\\
&&  \qquad \times \Big\{ K_{n}(l_+,\bar v_0;v_0,\vec w\,)\,
\rho_\pi(|\bar{v}_+|, \vec w-\vec l\;)\, \Big[  \Theta(+\bar{v}_+
)-\Theta(k_F- | \vec l\,| ) \Big]
\nonumber\\
&&  \qquad\;+ \,K_{n}(l_-,\bar v_{0};v_0,\vec
w\,)\,\rho_\pi(|\bar{v}_-|, \vec w-\vec l\;)\; \Theta(-\bar{v}_-
)\, \Big\} \,,
\nonumber\\
&& l_\pm^\mu = (\pm \,\sqrt{m_N^2+\vec l\,^2}, \vec l \;)\,,
\qquad \bar{v}_\pm = \bar{v}_0  \mp \sqrt{m_N^2+\vec l\,^2} \,,
\label{def-loop-J_n}
\end{eqnarray}
where the explicit form of the kernels $K_n$ as well as of the counter loops $J_n^C(v_0,\vec w)$
are recalled in \cite{Lutz:Korpa:Moeller:2007}. The kernels are invariant functions of
the 4-vectors $l_\mu, v_\mu, \bar v_\mu$ and $u_\mu$.
The spectral density of the pion, $\rho_\pi (\omega , \vec q\,)$, is
\begin{eqnarray}
&& \rho_\pi (\omega, \vec q\,) = -\Im \,\frac{1}{\omega^2-\vec q\,^2-m_\pi^2
-\Pi(\omega, \vec q\,)} \qquad {\rm for} \;\quad  \omega >0\,,
\nonumber\\
&& \rho_\pi (-\omega, \vec q\,) = -\rho_\pi (\omega, \vec q\,)\,.
\label{def-pion-spectral}
\end{eqnarray}

\section{Isobar self energy in the presence of vertex corrections}

\begin{figure}[b]
\begin{center}
\begin{eqnarray}
\Sigma^{\mu\nu}_{\Delta}=\parbox{14\unitlength}{\vspace{-0.8cm}\includegraphics[scale=0.6]{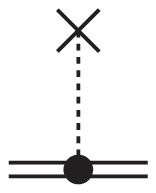}}\, +\,
\parbox{45\unitlength}{\vspace{-0.3cm}\includegraphics[scale=0.6]{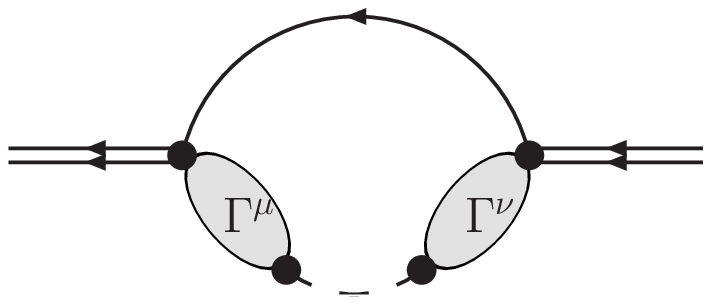}}\, +\,
\parbox{45\unitlength}{\vspace{-0.7cm}\includegraphics[scale=0.6]{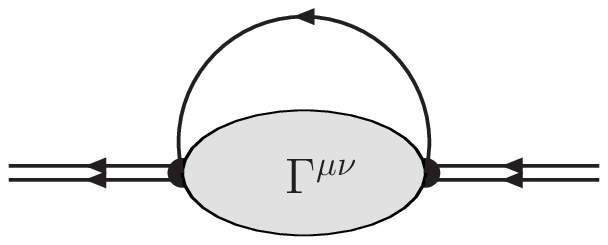}}\,\nonumber
\end{eqnarray}
\end{center}
\caption{Isobar self energy in the presence of short rage correlations. The solid line shows a nucleon
propagator in the presence of mean fields. The dashed line represents a dressed pion propagator.  }
\label{fig:3}
\end{figure}

The evaluation of the loop functions in the presence of
vertex corrections is particularly challenging due to their
complicated ultraviolet behavior. In Fig. 3 the two types of contributions are
depicted graphically in terms of vertex functions to be specified
below. It is useful to identify
a set of master loop functions, in terms of which the full loop
matrix can be constructed. The latter are renormalized applying the
scheme introduced in the previous section. The proper
generalization of (\ref{def-loop-J_n}) is readily worked out.  The pion spectral function
is distorted by vertex correction functions leading to effective spectral densities, which we denote with
$\rho_{ab}(\omega, \vec q\,)$. For a given spectral distribution we introduce
\begin{eqnarray}
&&\Delta J_{ab,n}(\bar v_0; v_0,\vec w\,) = \int
\frac{d\,^3 l}{2\,(2\, \pi)^3}\, \,\Big(m_N^2+\vec l\,^2
\Big)^{-\frac{1}{2}}\,
\nonumber\\
&&  \times \Big\{ K_{n}^{}(l_+,\bar v_0;  v_0, \vec w)\,
\rho_{ab}(|\bar v_+|, \vec w-\vec l\;)\, \Big[ \Theta(+\bar v_+
)-\left(\frac{\bar v_+}{|\bar v_+|}\right)^{a+b}\Theta(k_F- | \vec
l\,| ) \Big]
\nonumber\\
&&  \;+ \left(\frac{\bar v_-}{|\bar
v_-|}\right)^{a+b}\,K_{n}^{}(l_-,\bar v_0; v_0, \vec
w)\,\rho_{ab}(|\bar v_-|, \vec w-\vec l\;)\; \Theta(-\bar v_-  )\,
\Big\} \,,
\nonumber\\
&& l_\pm^\mu = (\pm \,\sqrt{m_N^2+\vec l\,^2}, \vec l \;)\,,
\qquad \bar v_\pm = \bar v_0  \mp \sqrt{m_N^2+\vec l\,^2} \,,
\label{def-loop-vertex}
\end{eqnarray}
where $n=0,...,12$.
The kernels $K_n(l,\bar v_0; v_0, \vec w)$ are
identical to those encountered in (\ref{def-loop-J_n}). They are listed in \cite{Lutz:Korpa:Moeller:2007}.
The real part of the
loop functions is computed applying the dispersion-integral representation
(\ref{disp-integral}). A corresponding generalization holds for the second term in (\ref{disp-integral}).

We identify the effective spectral distributions, $\rho_{ab}(\omega, \vec q\,)$ as
implied by the diagrams of Fig. 3. The vertex vector and tensor may be decomposed into invariants
\begin{eqnarray}
&& \Gamma^\mu (q,u) = q^\mu\,\Gamma_1(q,u)+u^\mu\,\Gamma_2(q,u) \,,
\nonumber\\
&&\Gamma^{\mu \nu} (q,u) = q^\mu\,q^\nu\Gamma_{11}(q,u)+q^\mu \,u^\nu\,\Gamma_{12}(q,u)
+u^\mu\,q^\nu\,\Gamma_{12}(q,u)
\nonumber\\
&& \qquad \qquad +\, u^\mu\,u^\nu\,\Gamma_{22}(q,u) + g^{\mu \nu}\,\Gamma_{00}(q,u)\,,
\label{}
\end{eqnarray}
in terms of which we introduce the spectral distributions
\begin{eqnarray}
&& \rho_{00}(\omega, \vec q\,) =
-\Im\,\Bigg(\Gamma_{00}(\omega,\vec q\,) \Bigg) \,,
\nonumber\\
&& \rho_{ab}(\omega, \vec q\,) =
-\Im\,\Bigg(\frac{\Gamma_a(\omega, \vec q\,)\,\Gamma_b(\omega, \vec q\,)}{\omega^2-\vec q\,^2-m_\pi^2
-\Pi_\pi(\omega, \vec q\,)} +\Gamma_{ab}(\omega,\vec q\,) \Bigg)\,.
\label{Aeff}
\end{eqnarray}

Applying the techniques introduced in \cite{Lutz:Migdal} it is straight forward to
express $\Gamma_{1}(q,u)$ and $\Gamma_{2}(q,u)$ in terms of the
longitudinal coupling matrix, $g^{(L)}$, and
the loop functions, $\Pi^{(L)}(q,u)$ of (\ref{def-decom-pi},\ref{def-matrix}). We obtain:
\begin{eqnarray}
&& \Gamma_1(q ,u) = \Big[1-g^{(L)}\,\Pi^{(L)}(q,u)\Big]^{-1}_{31} +
\Big[1-g^{(L)}\,\Pi^{(L)}(q,u)\Big]^{-1}_{33} \nonumber\\
&& \qquad  + \frac{q\cdot u }{\sqrt{q^2-(q\cdot u
)^2}}\,\left(\Big[1-g^{(L)}\,\Pi^{(L)}(q,u)\Big]^{-1}_{41} +
\Big[1-g^{(L)}\,\Pi^{(L)}(q,u)\Big]^{-1}_{43} \right) \,,
\nonumber\\
&& \Gamma_2(q ,u) = -\frac{q^2 }{\sqrt{q^2-(q\cdot u
)^2}}\,\Bigg(\Big[1-g^{(L)}\,\Pi^{(L)}(q,u)\Big]^{-1}_{41}
\nonumber\\
&& \qquad \qquad \qquad +
\Big[1-g^{(L)}\,\Pi^{(L)}(q,u)\Big]^{-1}_{43} \Bigg)\! \,.
 \label{def-Gamma}
\end{eqnarray}
The matrix $\Gamma_{ab}(q,u)$ probes longitudinal and transverse
correlations. As an extension of (\ref{def-matrix}) we introduce a transverse coupling and loop matrix $g^{(T)}$ and
$\Pi^{(T)}(q,u)$. We write
\begin{eqnarray}
&& g^{(T)}= \left(
\begin{array}{cc}
g_{11}' & g_{12}' \\
g_{21}' & g_{22}'
\end{array}
\right)
\,, \qquad
\Pi^{(T)}(q,u) =\left(
\begin{array}{cc}
\Pi_{\,T}^{(Nh)}(q,u) & 0 \\
0 &\Pi_{\,T}^{(\Delta h)}(q,u)
\end{array}
\right) \,.
\label{def-transverse}
\end{eqnarray}
We derive explicit forms of the tensor vertex
\begin{eqnarray}
&&\Gamma_{11}(q,u)= \frac{1}{q^2}\,\Big( \chi^{(L)}_{33}  +\frac{q \cdot u}{\sqrt{q^2-(q \cdot u)^2}}\,(\chi^{(L)}_{34}
+\chi^{(L)}_{43})
\nonumber\\
&& \qquad \qquad +\frac{(q \cdot u)^2}{q^2-(q \cdot u)^2}\,\chi^{(L)}_{44}
-\frac{q^2}{q^2-(q \cdot u)^2}\,\chi^{(T)}_{22}\Big)\,,
\nonumber\\
&&\Gamma_{12}(q,u)=\Gamma_{21}(q,u)= -\frac{1}{\sqrt{q^2-(q\cdot u)^2}}\,\chi^{(L)}_{34}-\frac{q \cdot u}{q^2-(q \cdot u)^2}\,\Big(\chi^{(L)}_{44}-\chi^{(T)}_{22}\Big)\,,
\nonumber\\
&&\Gamma_{22}(q,u)=\frac{q^2}{q^2-(q \cdot u)^2}\,\Big(\chi^{(L)}_{44}-\chi^{(T)}_{22}\Big)\,,
\qquad \quad \Gamma_{00}(q,u) = \chi^{(T)}_{22} \,,
\end{eqnarray}
in terms of the longitudinal and transverse correlation functions
\begin{eqnarray}
&& \chi^{(L,T)}(q,u) = \Big[1-g^{(L,T)}\,\Pi^{(L,T)}(q,u)\Big]^{-1}\,g^{(L,T)} \,.
\end{eqnarray}

In the course of deriving the representation
(\ref{def-loop-vertex}) we made use of the following  properties of the
correlation functions
\begin{eqnarray}
&&\Pi^{(L)}_{ij}(-\omega, \vec q\,)= (-1)^{i+j}\,\Pi^{(L)}_{ij}(+\omega , \vec q\,) \,,
\\
&& \Pi^{(T)}_{ij}(-\omega, \vec q\,)= (-1)^{i+j}\,\Pi^{(T)}_{ij}(+\omega , \vec q\,) \,,
\nonumber\\
&&\Gamma_a(-\omega, \vec q\,) =(-1)^{a+1}\,\Gamma_a(+\omega, \vec q\,)\,, \quad
\Gamma_{ab}(-\omega, \vec q\,) =(-1)^{a+b}\,\Gamma_{ab}(+\omega, \vec q\,)  \,. \nonumber
\label{property-vertex}
\end{eqnarray}

It is left to specify the isobar self energy in terms of the generic loop functions defined by
(\ref{def-loop-vertex}). In a first step a matrix of loop functions,
$J^{(p,q)}_{ab,[ij]}(v,u)$, is constructed in terms
of $J^{}_{ab,n}(v,u)$ as detailed in \cite{Lutz:Korpa:Moeller:2007}. The latter correspond to the projector algebra of section 4.
The evaluation of the self energy is analogous to the computation of section 4 with the slight complication that
the effective vertex develops additional structures $q_\mu\,u_\mu+u_\mu\,q_\nu, u_\mu u_\nu $ and $g_{\mu \nu}$.
The loops $J^{(p,q)}_{11,[ij]}(v,u)$, which are implied by the structure $q_\mu \,q_\nu$, contribute
like the previous loops $J^{(p,q)}_{[ij]}(v,u)$ in (\ref{def-sigma-identify}). The implication of the
remaining loop functions is readily worked out upon the application of the useful identities
\begin{eqnarray}
&&u^\mu = -i\,\sqrt{\frac{2}{3}}\,\sqrt{\frac{(v\cdot u)^2}{v^2}-1}\,\Big\{\bar P_{[17]}^\mu+\bar P_{[28]}^\mu
-\frac{1}{\sqrt{2}}\,(\bar P_{[14]}^\mu+\bar P_{[23]}^\mu) \Big\}
\nonumber\\
&& \qquad +\, \frac{v \cdot u}{\sqrt{v^2}}\,
(\bar P_{[15]}^\mu+\bar P_{[26]}^\mu)
\nonumber\\
&& \quad\; =-i\,\sqrt{\frac{2}{3}}\,\sqrt{\frac{(v\cdot u)^2}{v^2}-1}\,\Big\{P_{[71]}^\mu+P_{[82]}^\mu
-\frac{1}{\sqrt{2}}\,(P_{[41]}^\mu+P_{[32]}^\mu) \Big\}
\nonumber\\
&& \qquad +\, \frac{v \cdot u}{\sqrt{v^2}}\,
(P_{[51]}^\mu+P_{[62]}^\mu) \,,
\nonumber\\ \nonumber\\
&& g^{\mu \nu} \,P_{[11]}= Q_{[11]}^{\mu \nu}
+P_{[44]}^{\mu \nu} +P_{[55]}^{\mu \nu} + P_{[77]}^{\mu \nu} \,,
\nonumber\\
&& g^{\mu \nu} \,P_{[22]}= Q_{[22]}^{\mu \nu}
+P_{[33]}^{\mu \nu} +P_{[66]}^{\mu \nu} + P_{[88]}^{\mu \nu} \,,
\nonumber\\
&& g^{\mu \nu} \,P_{[12]}= Q_{[12]}^{\mu \nu}
-{\textstyle{1\over 3}}\,P_{[43]}^{\mu \nu} +P_{[56]}^{\mu \nu}+
{\textstyle{1\over 3}}\,P_{[78]}^{\mu \nu} -{\textstyle{\sqrt{8}\over 3}}\,(P_{[73]}^{\mu \nu}+P_{[48]}^{\mu \nu})\,,
\nonumber\\
&& g^{\mu \nu} \,P_{[21]}= Q_{[21]}^{\mu \nu}
-{\textstyle{1\over 3}}\,P_{[34]}^{\mu \nu} +P_{[65]}^{\mu \nu}+
{\textstyle{1\over 3}}\,P_{[87]}^{\mu \nu}
-{\textstyle{\sqrt{8}\over 3}}\,(P_{[84]}^{\mu \nu}+P_{[37]}^{\mu \nu})\,.
\label{u-proj}
\end{eqnarray}
It is now straight forward to write down the self
energies, $\Sigma_{[ij]}^{(p,q)}(v,u)$. It holds
\begin{eqnarray}
&&\Sigma_{[ij]}^{(q)}(v,u) = - \frac{f_\Delta^2}{m_\pi^2}\,\Big\{J^{(q)}_{11,[ij]}(v,u)+J^{(p)}_{00,[ij]}(v,u) \Big\}\,,
\nonumber\\
&&\Sigma_{[ij]}^{(p)}(v,u) = - \frac{f_\Delta^2}{m_\pi^2}\,\Big\{
 J^{(p)}_{11,[ij]}(v,u) + \sum_{a,b\,=1}^2\,J^{(p)}_{22,[ab]}(v,u)\,c_{ai}(v,u)\,c_{bj}(v,u)
\nonumber\\
&& \qquad \qquad \quad   + \sum_{a=1}^2\,\Big(  J^{(p)}_{12,[ia]}(v,u)\,c_{aj}(v,u) +
 J^{(p)}_{21,[aj]}(v,u)\,c_{ai}(v,u)
 \nonumber\\
&& \qquad \qquad \quad   + \sum_{a,b=1}^2\,J^{(p)}_{00,[ab]}(v,u)\,c^{(ab)}_{[ij]}(v,u)
\Big\}\,,
\nonumber\\
&& c_{aj}(v,u)  = \frac{ v \cdot u}{\sqrt{v^2}}\, \delta_{4+a,j}
-i\,\sqrt{\frac{2}{3}}\,\sqrt{\frac{(v\cdot u)^2}{v^2}-1}\,(\delta_{6+a,j}- \frac{1}{\sqrt{2}}\,\delta_{5-a,j})\,,
\nonumber\\ \nonumber\\
&& c^{(ab)}_{[ij]}(v,u) = \delta_{a1}\,\delta_{b2} \,
\Big({\textstyle{1\over 3}}\,(\delta_{i7}\,\delta_{j8}-\delta_{i4}\,\delta_{j3})+\delta_{i5}\,\delta_{j6}
-{\textstyle{\sqrt{8}\over 3}}\,(\delta_{i7}\,\delta_{j3}+\delta_{i4}\,\delta_{j8} )\Big)
\nonumber\\
&& \qquad \qquad \; +\,\delta_{ab}\,\delta_{ij}\,\Big( \delta_{i,5-a}+\delta_{i,4+a}+\delta_{i,6+a}\Big)
\nonumber\\
&& \qquad \qquad \; +\,\delta_{a2}\,\delta_{b1} \,
\Big({\textstyle{1\over 3}}\,(\delta_{i8}\,\delta_{j7}-\delta_{i3}\,\delta_{j4})+\delta_{i6}\,\delta_{j5}
-{\textstyle{\sqrt{8}\over 3}}\,(\delta_{i3}\,\delta_{j7}+\delta_{i8}\,\delta_{j4} )\Big)  \,.
\label{isobar-self-ex}
\end{eqnarray}

\clearpage

\section{Numerical results and discussions}

We present and discuss numerical simulations of the pion and isobar spectral distributions at
nuclear saturation density  with the Fermi momentum $k_F=270\,$MeV. The results depend on
a number of parameters appearing in the developed covariant and self consistent approach. These are first of all the scalar and vector
mean-field shifts of the delta, $\Sigma^{\Delta}_{S}$ and $\Sigma^{\Delta}_{V}$, as well as the Migdal parameters
$g'_{11}$, $g'_{12}$ and $g'_{22}$. One should also consider medium induced changes in the
coupling constants $f_\Delta$ and $f_N$, although it is usually assumed that for nuclear
densities they do not depart significantly from their vacuum values. The nucleon mean-field
parameters $\Sigma_N^S$ and $\Sigma_V^S$, which model nuclear saturation and binding effects,
are also prone to variations in different models. We use the values $\Sigma_S^N = 350$ MeV and $\Sigma_V^N = 290$ MeV
also assumed in \cite{Lutz:Korpa:Moeller:2007,Riek:Lutz:Korpa:2008}.

\begin{figure}[b]
\begin{center}
\hskip-1.0cm
\parbox{12.cm}{
\parbox{5.8cm}{\includegraphics[scale=0.65]{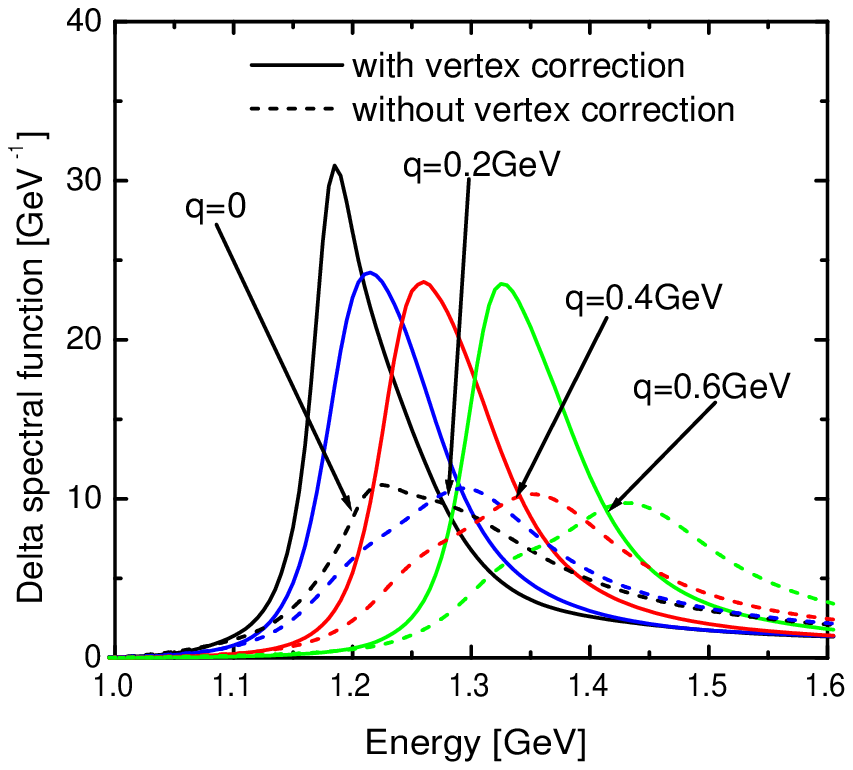}}\parbox{5.8cm}{\includegraphics[scale=0.65]{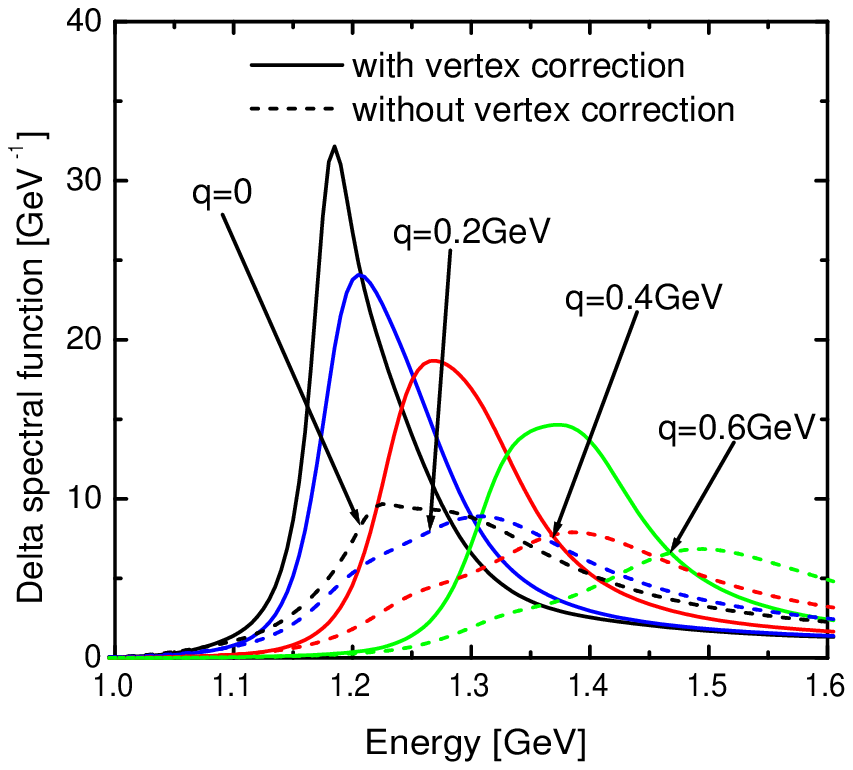}} \\
\vspace*{-0.3cm}
\hspace*{-0.6cm}
\parbox{5.8cm}{\includegraphics[scale=0.65]{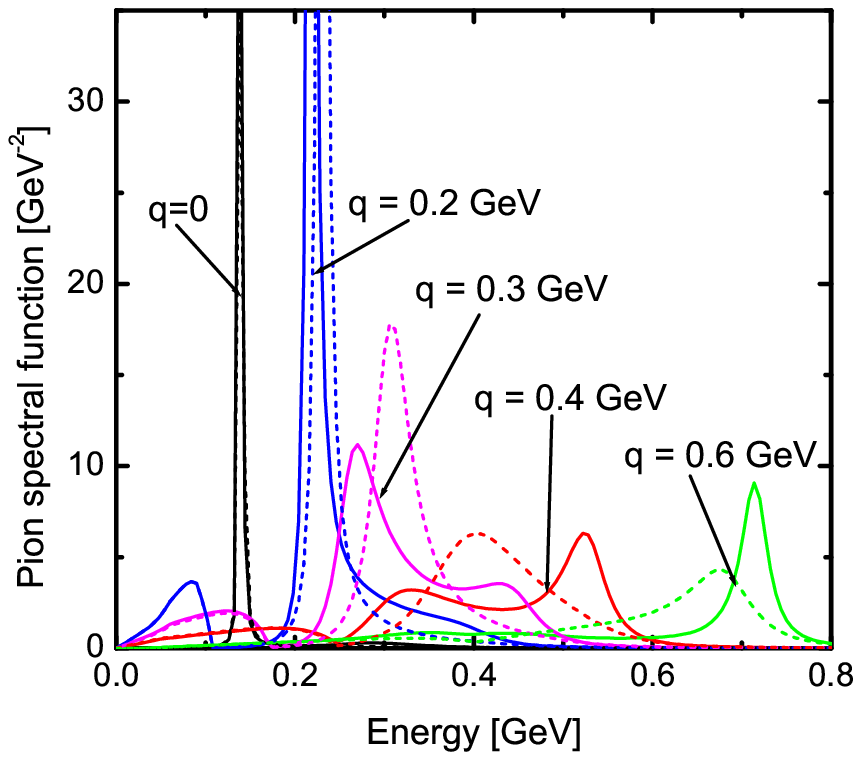}}\parbox{5.8cm}{\includegraphics[scale=0.65]{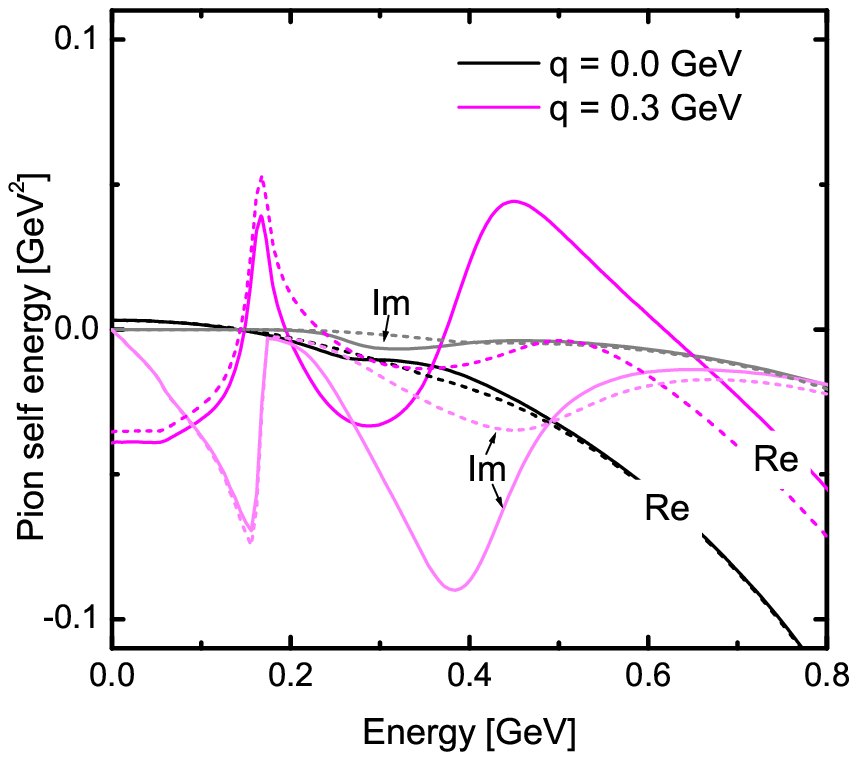}}}
\hspace*{1.0cm}\parbox{3.cm}{\vspace*{-1.3cm}
\caption{Results for the delta propagator with (solid lines) and without (dashed lines) the $\pi\rm{N}\Delta$ vertex correction (upper
figures). We show the $S_{[77]}^{(p)}$ component (left figure) and the $S_{[11]}^{(q)}$ component (right figure)
for different momenta. The pion spectral function (left figure) and self-energy (right figure)
are shown on the lower figures ($q$ is the pion momentum). The parameters of (\ref{central-set}) are used together with our reduced
value for $f_\Delta$.}
\label{fig:4} }
\end{center}
\vskip-0.7cm
\end{figure}

\begin{figure}[t]
\vskip-0.1cm
\begin{center}
\hskip-1.cm
\parbox{12.cm}{\includegraphics[scale=1.3]{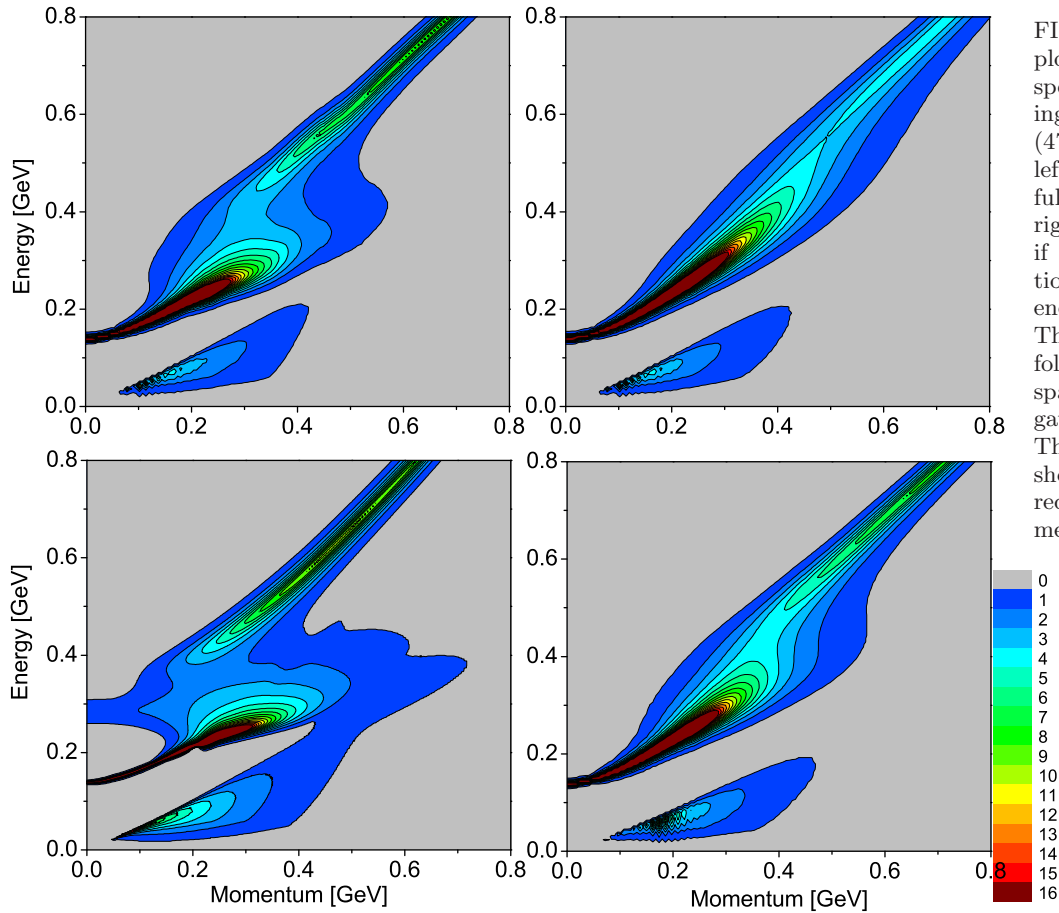}}
\hspace*{2.3cm}\parbox{3.0cm}{\vspace*{-5.cm}
\caption{Contour plots of the pion spectral function using the parameters (\ref{central-set}). The
upper left figure gives our full result, the upper right figure follows if the vertex correction in
the delta self-energy are neglected.  The lower left figure follows if the free-space isobar
propagator is used in (\ref{full-result}).  The lower right figure shows the impact of reducing the nucleon mean fields.}
\label{fig:5} }
\end{center}
\vskip-0.9cm
\end{figure}

As a guide we consider the values of the above parameters used in earlier computations,
but having in mind their scheme dependence. We focus on variations around a parameter set
which has been shown to lead to a reproduction of the nuclear photoabsorption cross-section in the delta
excitation region \cite{Riek:Lutz:Korpa:2008}. The latter study builds on the self-consistent approach developed
in this work. It is the first work that considers photabsorption in the presence of short-range correlation effects
in the $\gamma \,\pi \,\pi$, $\gamma \,N \,\Delta$, $\gamma\,\pi \,N\,\Delta $, $\pi\,N\,\Delta $ and $\pi \,N \,N$ vertices.
Electromagnetic gauge invariance is kept as a consequence of a series of Ward identities obeyed in the computation. In particular
the interference of the in-medium s-channel isobar exchange and the t-channel in-medium pion exchange is considered.
We refer to the details of that work which provides the following parameter set:
\begin{eqnarray}
&&\Sigma^{\Delta}_{S}=-0.25\,\rm{GeV}\,,\qquad\Sigma^{\Delta}_{V}=-0.11\,\rm{GeV}\,,\qquad
g'_{11}=1.0\,,\qquad g'_{12}=0.4\,,\qquad g'_{22}=0.4\,,
\label{central-set}
\end{eqnarray}
together with an in-medium reduction of the $f_\Delta$ coupling by 15\% but an unchanged value for $f_N$.
According to \cite{Oset:Weise:76} the in-medium reduction of the $f_N$ coupling is quite
small (less than 6\% at saturation density). This is in line with our finding that the photoabsorption data does not
require any in-medium change of $f_N$.  The values of the  Migdal parameters in (\ref{central-set}) is within range of
the various sets used in the literature. Though in the recent work by Hees and Rapp \cite{Hees:Rapp}  large values
of $g'_{12}$ and $g'_{22}$ are excluded in their non-relativistic scheme, this is not the case in our more microscopic and relativistic approach.
Values for $g'_{12}$ and $g'_{22} $ as large as in (\ref{central-set}) generate a width for the isobar in \cite{Hees:Rapp}
that would be incompatible with the photoabsorption data as computed in \cite{Hees:Rapp}. However, it is reasonable to expect that
such a condition is altered by a possible in-medium reduction of $f_\Delta$.

We recall from  \cite{Oset:Salcedo,Riek:Lutz:Korpa:2008} that the actual position of the photoabsorption peak is a subtle effect of short-range
correlation effects and the in-medium isobar properties. The peak of the isobar spectral distribution does not translate directly
into the maximum of the photoabsorption cross section.  The pion and isobar properties as implied by (\ref{central-set})
are shown in Fig. \ref{fig:4} for nuclear saturation densities by solid lines: at zero momentum the
isobar receives an attractive mass shift of about 50 MeV. A value amazingly close to the range obtained in \cite{Oset:Salcedo} but in stark
contrast to the small and repulsive mass shift obtained recently in \cite{Hees:Rapp}. For the isobar we restrict the discussion to the two main components because they dominate the resonance region. Please note however, that the proper inclusion of all other components
is essential to ensure the cancelation of kinematical singularities on the light-cone.
We observe a significant splitting of the p- and q-space modes at nonzero momentum. The medium
effects are stronger for the q-space (helicity 3/2) than they are for the p-space (helicity 1/2),
where we obtain a less pronounced broadening and smaller shift in the position of the peak at larger
momentum. Note that the nuclear photoabsorption data probe dominantly the
helicity 3/2 mode. These finding are in qualitative agreement with the results of \cite{Oset:Salcedo} that were based on a perturbative and
non-relativistic many-body approach. It should be pointed out, however, that the pion spectral function corresponding to the approach of \cite{Oset:Salcedo}
differs decisively from the one predicted by our approach. Though a direct comparison is difficult, since Oset et al did not provide figures for the
pion spectral function, an indirect comparison may be possible. We take the more recent work of Ramos and Oset \cite{Ramos:Oset:2000}, which provides
explicit results for the pion spectral distribution.  The strength in the soft pion modes as shown in Fig. \ref{fig:3} is much suppressed as
compared to an in-medium pion considered realistic in \cite{Ramos:Oset:2000}. Also a comparison of our pion spectral function in Fig. \ref{fig:3}
with other recent results \cite{Korpa:Lutz:04,Post:Leupold:Mosel,Hees:Rapp} show significant and systematic differences at small and intermediate momenta.

\begin{figure}[t]
\vskip-0.1cm
\begin{center}
\hskip-1.cm
\parbox{11.cm}{
\parbox{5.5cm}{\includegraphics[scale=0.6]{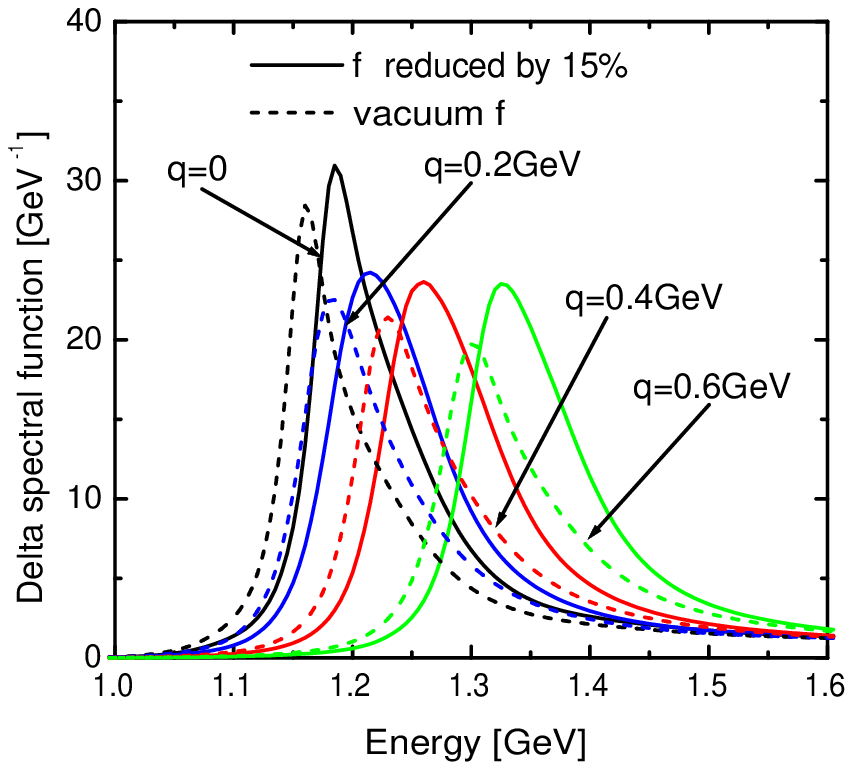}}\parbox{5.5cm}{\includegraphics[scale=0.6]{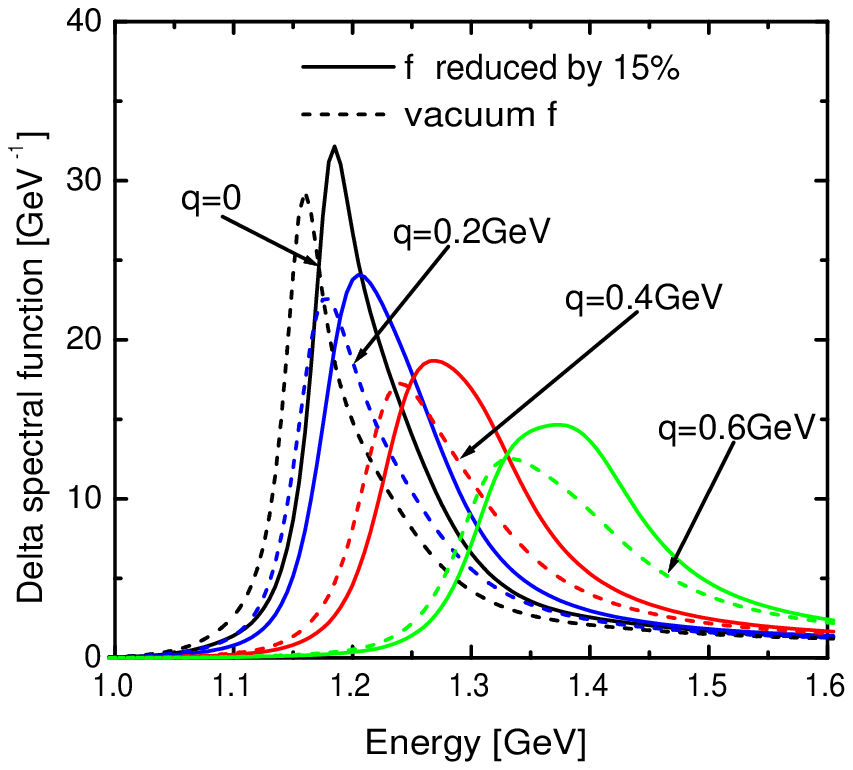}} \\
\vspace*{-0.3cm}
\hspace*{-0.5cm}
\parbox{5.5cm}{\includegraphics[scale=0.6]{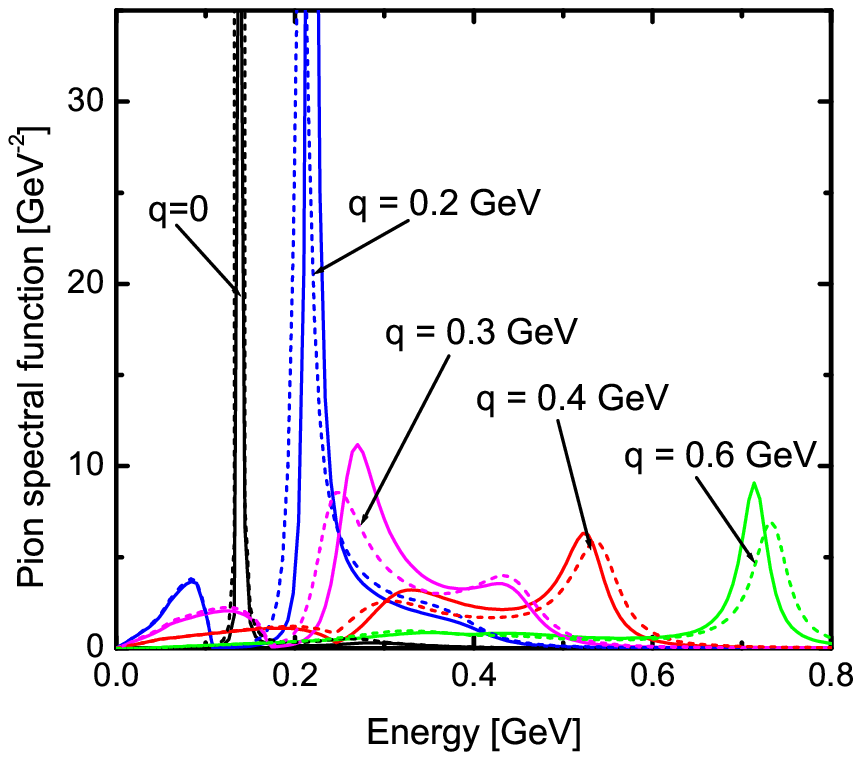}}\parbox{5.5cm}{\includegraphics[scale=0.6]{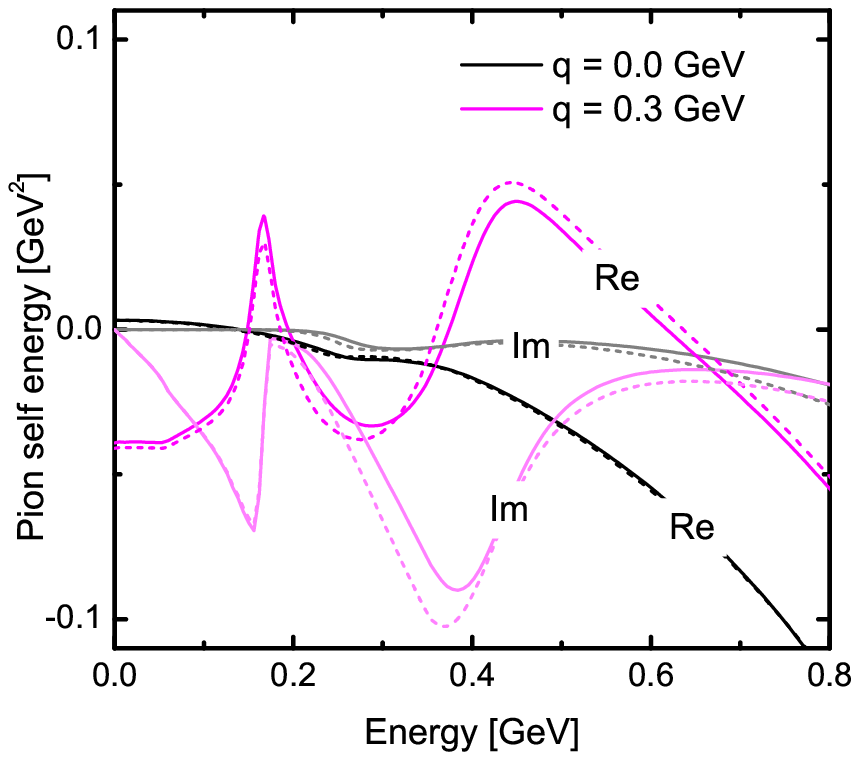}}}
\hspace*{1.cm}\parbox{3.0cm}{\vspace*{-1.7cm}
\caption{Same as Fig.\ \ref{fig:4}, but varying the  $\pi N \Delta $ coupling constant $f_\Delta$. The solid lines
correspond to the $15 \%$ reduction of $f_\Delta$, the dashed ones to its free-space value. }
\label{fig:7} }
\end{center}
\vskip-0.8cm
\end{figure}

\begin{figure}[b]
\vskip-0.4cm
\begin{center}
\hskip-1.cm
\parbox{11.cm}{
\parbox{5.5cm}{\includegraphics[scale=0.6]{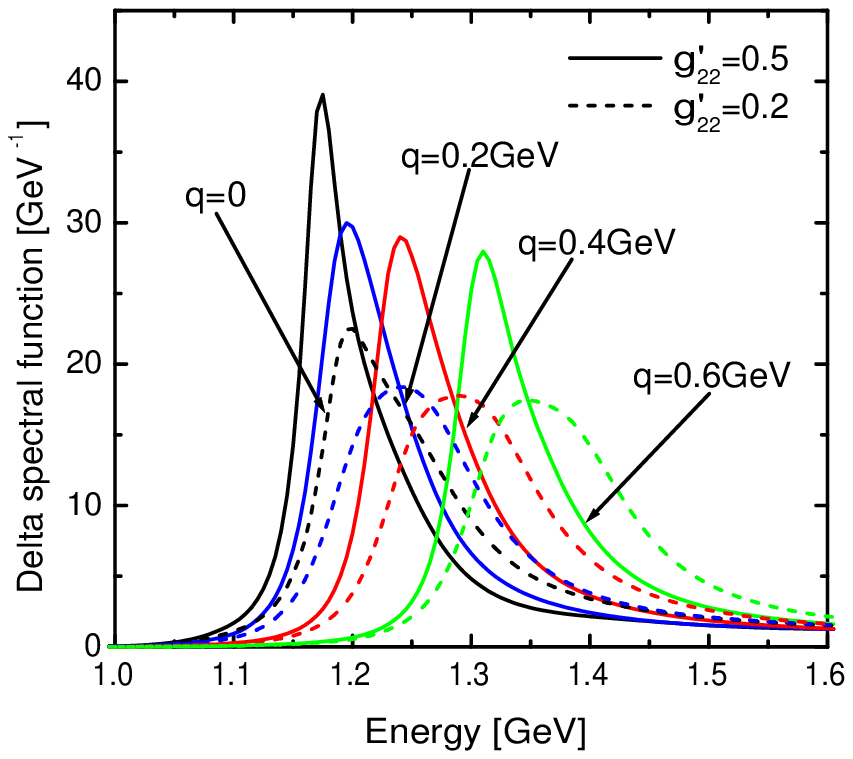}}\parbox{5.5cm}{\includegraphics[scale=0.6]{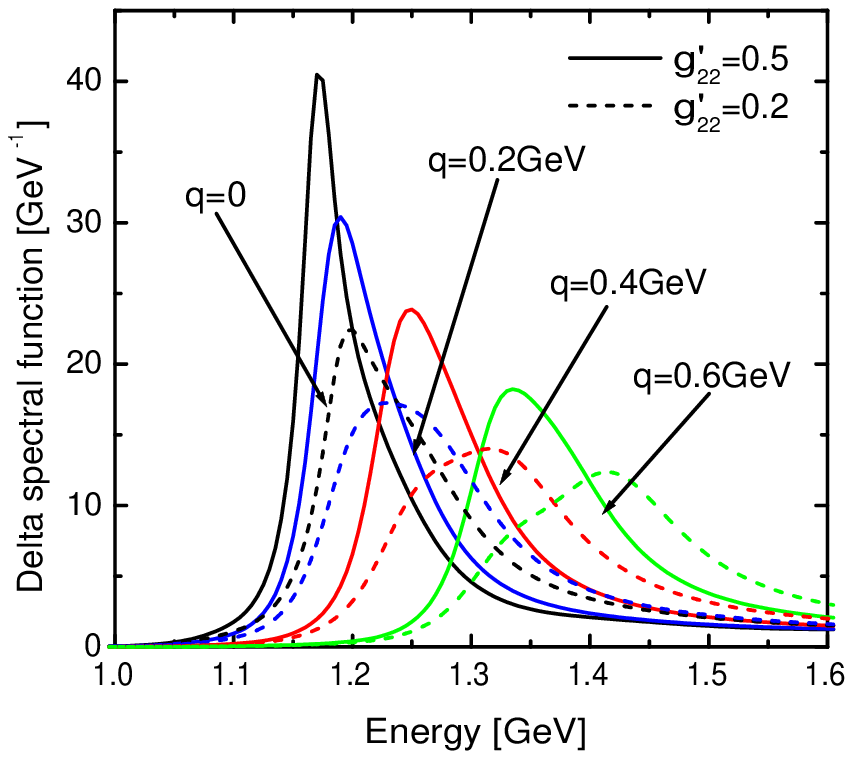}} \\
\vspace*{-0.3cm}
\hspace*{-0.5cm}
\parbox{5.5cm}{\includegraphics[scale=0.6]{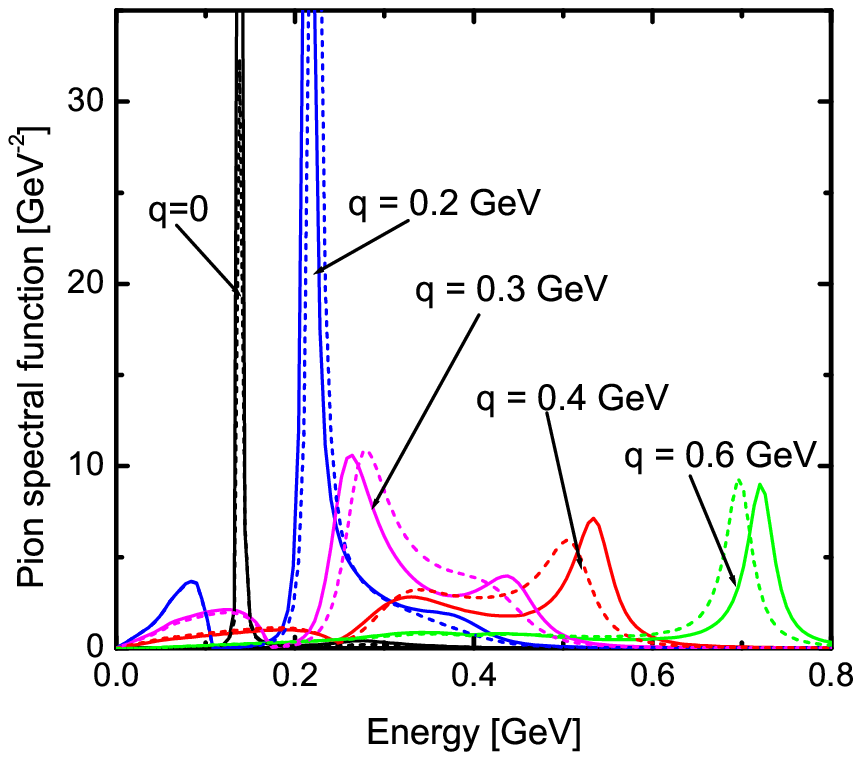}}\parbox{5.5cm}{\includegraphics[scale=0.6]{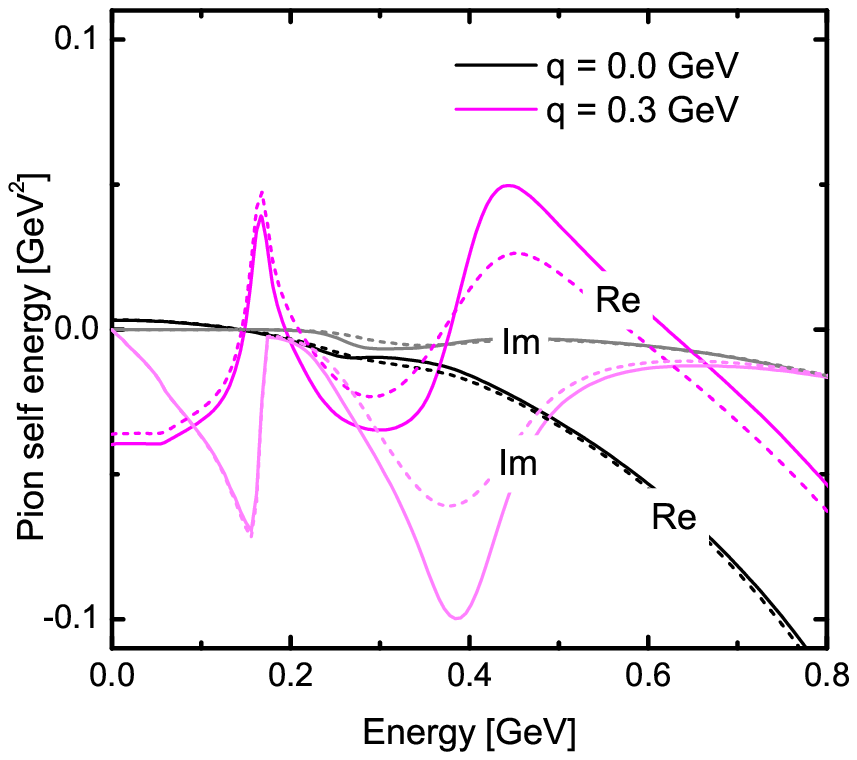}}}
\hspace*{1.cm}\parbox{3.0cm}{\vspace*{-1.7cm}
\caption{Same as Fig.\ \ref{fig:4} but for different values of the $g'_{22}$
parameter. The solid lines correspond to the $g_{22}'=0.5$, the dashed ones to $g_{22}'=0.2$. }}
\end{center}
\vskip-0.8cm
\label{fig:8}
\end{figure}

Before we discuss a variation of parameters around the central values (\ref{central-set}) we examine
the effect of various approximations all based on the parameter set (\ref{central-set}). First we consider the effect of
neglecting short-range correlation effects in the isobar self energy as done in \cite{Xia:Siemens:Soyeur,Korpa:Lutz:04,Post:Leupold:Mosel}.
In Figs. \ref{fig:4} and \ref{fig:5} the quality of such an approximation is scrutinized. Though the Migdal parameters enter in a
decisive manner in the computation of the pion-self energy via (\ref{full-result}), the isobar properties are a functional of the
pion self energy only. As studied in great detail in the previous works \cite{Xia:Siemens:Soyeur,Korpa:Lutz:04,Post:Leupold:Mosel}, the self
consistent treatment of the pion and isobar properties is an important and significant effect even in the absence of vertex corrections for the isobar.
The upper left panel of Fig. \ref{fig:5} shows the contour lines of the pion spectral function as obtained in the fully self consistent computation. If one neglects the $\pi\rm{N}\Delta$ vertex correction in the isobar self-energy as discussed above, the contour lines in the upper right-hand panel arise. A more
quantitative illustration is offered by a comparison of the solid and dashed lines in Fig. \ref{fig:4}.
The Figs. \ref{fig:4} and \ref{fig:5}  document the importance of the vertex corrections in the isobar self energy.  Most significant
are the effects on the isobar spectral distribution as shown by the solid and dashed lines of Fig. \ref{fig:4}. The consistent consideration
of short range correlation effects leads to a significant attractive mass shift and a reduction of the width for the isobar.
It is interesting to observe that it appears well justified to treat the vertex contributions in the isobar self energy in perturbation theory.
We find that the evaluation of the vertex bubbles of Fig. \ref{fig:3} with a free-space isobar propagator leads to results that can barely
be discriminated from our full results. Recall, however, that a corresponding attempt for the short-range bubbles in the pion self energy would fail
miserably. This is illustrated by the lower left-hand panel of Fig. \ref{fig:5}, where the pion spectral function is shown as it is implied by
the free-space isobar together with the Migdal parameters of (\ref{central-set}) and our in-medium value for $f_\Delta$.
In particular the width of the low-momentum main pion mode would be underestimated.

We now turn to a variation of the parameter set. In the lower right-hand panel Fig. \ref{fig:5} the effect of
using smaller scalar and vector mean fields for the nucleons is illustrated. The contour lines were obtained with
$\Sigma_S^N = 175$ MeV and $\Sigma_V = 115$ MeV. In Fig. \ref{fig:7} the variation of the value chosen for the $\pi\rm{N}\Delta$
coupling constant $f_\Delta$ is investigated. The reason for considering the departure from the
vacuum value is that a detailed study \cite{Riek:Lutz:Korpa:2008} of nuclear photoabsorption
strongly favors such a change, more precisely a reduction of the $f_\Delta$ coupling by
about $(10-15)\%$ at nucleon densities close to saturation. As expected a reduction of the coupling by
leads to a reduction of the isobar width. In the pion spectral function we can, at least for intermediate momenta,
distinguish three branches.
These are the main pion mode as well as the particle-hole and $\Delta$-hole excitation.
At about 0.3 GeV momentum we observe the level crossing between the main pion mode and the isobar-hole
excitation. Decreasing $f_\Delta$ reduces the strength of the
isobar-hole branch and in addition due to the narrower isobar that mode becomes better visible.

Next we study the influence of Migdal's $g'_{22}$  parameter. Varying its value from 0.2 to 0.5
we arrive at the results shown in Fig. \ref{fig:8}. The effect of changing $g'_{22}$ is subtle
since it influences the dressing of the isobar through the $\pi\rm{N}\Delta$ vertex correction and also
the pion self energy by affecting the isobar-hole loop contribution. Increasing the value of
$g'_{22}$ softens the isobar and decreases its width, which compensates in part the reduction of the
isobar-hole-loop contribution to the pion self energy. All together the resulting change in the pion
spectral function is modest. We note that a variation of $g'_{12}$ is quite similar to that of $g'_{22}$.
A variation of $g'_{11}$ just affects the nucleon-hole contribution. Lowering $g'_{11}$
makes the nucleon-hole branch of the pion larger, which in turn somewhat increases the isobar broadening.

\begin{figure}[t]
\vskip-0.1cm
\begin{center}
\hskip-1.cm
\parbox{11.cm}{
\parbox{5.5cm}{\includegraphics[scale=0.6]{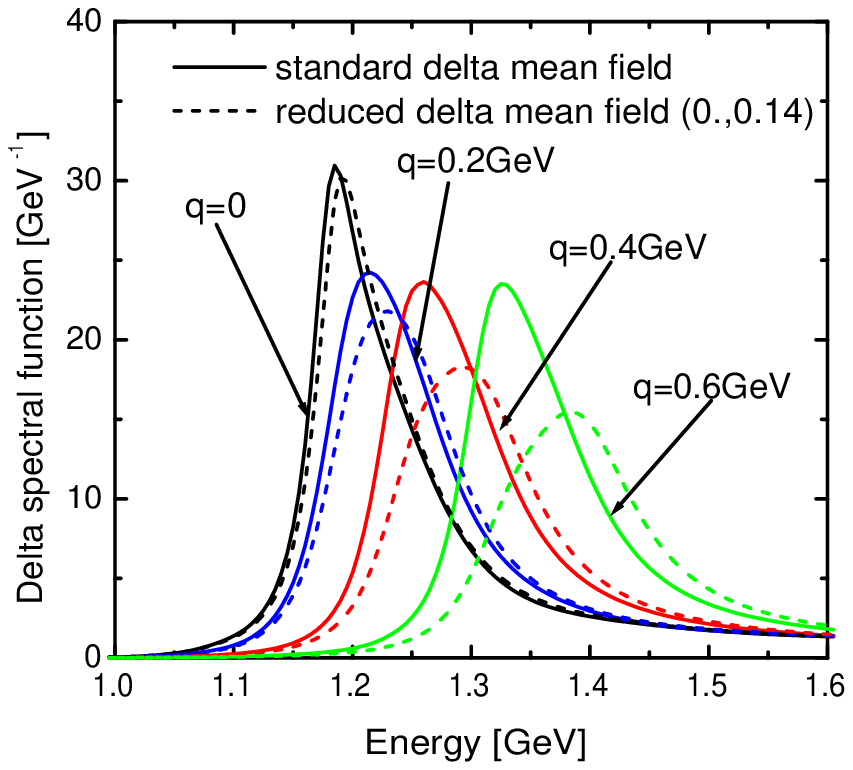}}\parbox{5.5cm}{\includegraphics[scale=0.6]{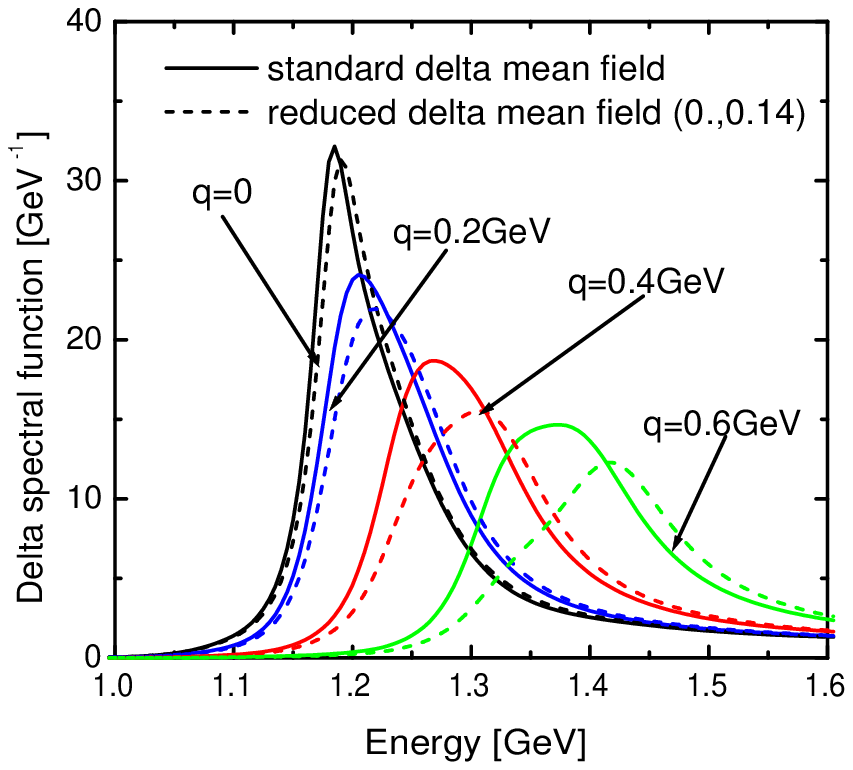}} \\
\vspace*{-0.3cm}
\hspace*{-0.5cm}
\parbox{5.5cm}{\includegraphics[scale=0.6]{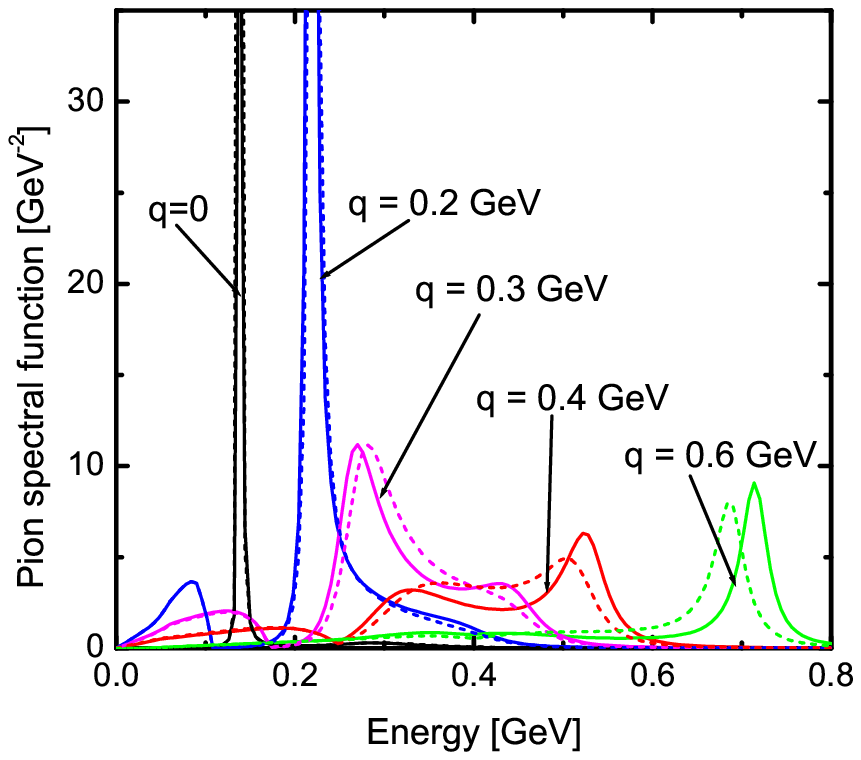}}\parbox{5.5cm}{\includegraphics[scale=0.6]{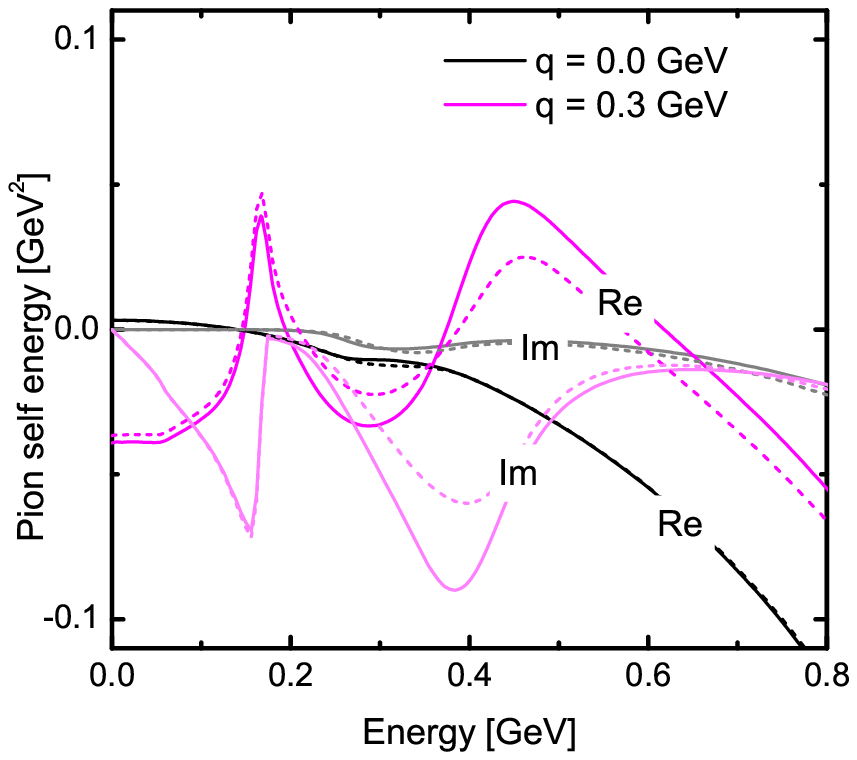}}}
\hspace*{1.cm}\parbox{3.0cm}{\vspace*{-1.7cm}
\caption{Same as Fig.\ \ref{fig:4} but for a variation of the delta mean-fields
$\Sigma^{\Delta}_{S}$ and $\Sigma^{\Delta}_{V}$. The solid lines correspond to the standard choice of (\ref{central-set}),
the dashed lines to the choice $\Sigma^{\Delta}_{S}=0.00$ GeV and $\Sigma^{\Delta}_{V}=0.14$ GeV.}}
\end{center}
\vskip-0.8cm
\label{fig:9}
\end{figure}

We conclude with a discussion of the influence of the isobar mean field parameters. Results are shown for two parameter sets,
which induce the same energy shift at zero momentum. Next to our standard set we use a set whose
scalar mean field is put to zero and the vector part provides the net repulsion of $0.14$ GeV at zero momentum as is implied
also by (\ref{central-set}). The effects can be found in Fig. 8. Without the scalar mean field we obtain
less attraction at nonzero momentum and in addition the width of the isobar is significantly increased at larger momenta.
This implies a smaller contribution of the isobar-hole state to the self energy of the pion as shown in the lower right-hand panel
of Fig. 8.

\section{Summary}

A detailed study of pion and isobar properties in cold nuclear was presented.  A fully relativistic and self-consistent
many-body approach  was developed that is applicable in the presence of Migdal's short range correlations effects.
Nuclear saturation and binding effects were modeled by scalar and vector mean fields for the nucleon.
The novel subtraction scheme, that was constructed recently by two of the authors and that avoids the
occurrence of kinematical singularities, was used. Unlike in previous studies no soft form factors for the $\pi N \Delta$
vertex were needed.  For the first time the $\pi\rm{N}\Delta$ vertex corrections as dictated by Migdal's short-range
interactions were considered in a relativistic and self consistent many-body approach. The latter were found to affect
the isobar and pion properties dramatically. Using realistic parameters sets we predict a downward shift of about
50 MeV for the $\Delta$ resonance at nuclear saturation density. The pionic soft modes are much less pronounced
than in previous studies.

Further studies are needed to consolidate our results. In particular an application to the
pion-nucleus problem and the pionic atom data set would be useful to further constrain the
parameter set. Our computation may be generalized to study effects of finite temperature.

\vskip0.3cm

{\bfseries{Acknowledgments}}
\vskip0.3cm

This research was supported in part by the Hungarian Research
Foundation (OTKA) grant 71989. M.F.M. L. acknowledges stimulating
discussions with R. Rapp  and D. Voskresensky.
C.L.K.\ would like to thank the G.S.I.\ (Darmstadt) and the
K.V.I.\ (Groningen) for the
kind hospitality. F. R. acknowledges useful
discussions with J. Knoll and thanks the FIAS (Frankfurt) for support.
\clearpage

\section*{Appendix A}

We present explicit representations for the nucleon- and isobar-hole loops
introduced in (\ref{nh-dh}) for the case of nuclear matter at
rest with $u_\mu =(1, \vec 0\,)$. The longitudinal \cite{Lutz:Migdal} and transverse
nucleon-hole loop functions are:
\begin{eqnarray}
&& \Pi^{(Nh)}_{ij}(\omega, \vec q\,) = \frac{f^2_{N}}{m^2_\pi}\, {\mathcal P }
\int_0^{k_F}\,\frac{d^3p}{2\,(p_0-\Sigma_V)\,(2\pi)^3}\,\frac{8\,K_{ij}^{(Nh)}}{
2\,p\cdot q+q^2+i\,\epsilon}
\nonumber\\
&& \quad +\frac{i\,f^2_{N}}{m^2_\pi}
\Im
\int_0^{k_F}\,\frac{d^3p}{2\,(p_0-\Sigma_V)\,(2\pi)^3}\,\frac{8\,K_{ij}^{(Nh)}\,\Theta
\Big( k_F-|\vec p+\vec q\,|\Big)}{ 2\,p\cdot q+q^2+i\,\epsilon
}\,\,\Theta \big(p_0+\omega \big)
\nonumber\\
&& \quad +(-1)^{i+j} (q_\mu \to -q_\mu )\,, \label{matrix-nh-a}
\end{eqnarray}
where $q_\mu =(\omega , \vec q\,)$, $p_0 = \sqrt{m_N^2+\vec
p\,^2}+\Sigma_V$ and
\begin{eqnarray}
&& K_{11}^{(Nh)} = 2\,m_N^2 \,,\qquad \quad
K_{12}^{(Nh)}=K_{21}^{(Nh)}=0 \,, \qquad
\nonumber\\
&& K^{(Nh)}_{22}= \frac{\omega^2 -\vec q\,^2}{\vec q\,^2}\,\Big(
2\, \vec p\,^2 + \omega \, (p_0-\Sigma_V) + \vec p \cdot \vec q \Big)+2\,m_N^2\,\frac{\omega^2}{\vec q^2}\,,
\nonumber\\
&& K_{\;T}^{(N h)} = 3\,m_N^2+\omega \, (p_0-\Sigma_V) - \vec p \cdot \vec q
- \frac{1}{2}\,\Big( K_{11}^{(N h)}+K_{22}^{(N h)}\Big)\,.
\label{matrix-nh-b}
\end{eqnarray}

For a bare isobar propagator, $S_0^{\mu \nu}(w)$ as given in (\ref{Dyson}), the
longitudinal isobar-hole loop functions were
computed already in \cite{Lutz:Migdal}. We present here longitudinal as well as
the transverse loop functions:
\begin{eqnarray}
&& \Pi^{(\Delta h)}_{ij}(\omega, \vec q\,) = \frac{4}{9}
\frac{f^2_{\Delta }}{m^2_\pi}
\int_0^{k_F}\,\frac{d^3p}{2\,(p_0-\Sigma_V)\,(2\pi)^3}\,\frac{8\,K_{ij}^{(\Delta
h)}\,\big(m_N\,m_\Delta +m_N^2+(p\cdot q )\big)}{ 2\,p\cdot
q+q^2-m_\Delta^2+m_N^2+i\,\epsilon } \nonumber\\
&& \qquad \qquad \quad \;+(-1)^{i+j} (q_\mu \to -q_\mu )\,, \nonumber\\
\nonumber\\
&& K_{11}^{(\Delta h )} = 1- \frac{(q^2+p\cdot
q)^2}{q^2\,m^2_\Delta} \,,\quad K_{22}^{(\Delta h )} = 1 +
\frac{(\omega \,|\vec p \,|\,\cos
(\vec q\,,\vec p\,) - |\vec q\,|\,p_0 )^2}{m_\Delta^2\,q^2} \,, \nonumber\\
&& K_{12}^{(\Delta h )} = K_{21}^{(\Delta h )} = i\,\frac{q^2+p\cdot q
}{q^2\,m_\Delta^2} \,\big( |\vec q\,|\,p_0-\omega \,|\vec p
\,|\,\cos (\vec q\,,\vec p\,) \big)
\nonumber\\
&& K_{\;T}^{(\Delta h )} = 2- \frac{(p+q)^2}{2\,m_\Delta^2}
- \frac{1}{2}\,\Big( K_{11}^{(\Delta h)}+K_{22}^{(\Delta h)}\Big) \,,
\label{matrix-dh}
\end{eqnarray}
where $q_\mu =(\omega , \vec q\,)$, $p_0 = \sqrt{m_N^2+\vec
p\,^2}+\Sigma_V$. Both representations (\ref{matrix-nh-a}, \ref{matrix-dh}) are
compatible with (\ref{constraint-polarization}). On the other hand, only
(\ref{matrix-nh-a}) is consistent with (\ref{bounded-assumption}). The asymptotic
behavior of the isobar-hole loop as given in (\ref{matrix-dh}) is at odds with the condition
(\ref{bounded-assumption}).

To derive the general results for the
isobar-hole loop functions it is advantageous to choose a representation slightly
different to (\ref{matrix-dh}). We write
\begin{eqnarray}
&&\Pi_{11}^{(\Delta h )} (\omega, \vec q\,) = \frac{1}{q^2}\,\Pi_1^{(\Delta h )} (\omega, \vec q\,)\,,
\nonumber\\
&&\Pi_{12}^{(\Delta h )} (\omega, \vec q\,) =
\frac{1}{\sqrt{q^2-(q\cdot u)^2}}\left(\frac{q \cdot u}{q^2}\,
\Pi_1^{(\Delta h )} (\omega, \vec q\,)- \Pi_2^{(\Delta h )} (\omega, \vec q\,) \right)\,,
\nonumber\\
&& \Pi_{22}^{(\Delta h )} (\omega, \vec q\,) =\frac{q \cdot u}{q^2-(q \cdot u)^2}\,\Bigg(
\frac{q \cdot u}{q^2}\,\Pi_{1}^{\Delta h} (\omega, \vec q\,)
-2\,\Pi_{2}^{(\Delta h )} (\omega, \vec q\,)
\nonumber\\
&& \qquad \qquad \;\;  +\, \frac{q^2}{q\cdot u}\,\Pi_3^{(\Delta h )} (\omega, \vec q\,) \Bigg)\,,
\nonumber\\
&& \Pi_{T}^{(\Delta h)} (\omega, \vec q\,) = \frac{1}{2}\,\left(
\Pi_4^{(\Delta h )} (\omega, \vec q\,)-\Pi_{11}^{(\Delta h )} (\omega, \vec q\,)-\Pi_{22}^{(\Delta h)}
(\omega, \vec q\,)\right)\,.
\label{rewrite-dh}
\end{eqnarray}
The merit of the representation (\ref{rewrite-dh}) lies in its simple realization of the constraint equations
(\ref{constraint-polarization}). The first condition
is satisfied for any functions $\Pi_i(\omega, \vec q\,)$ that are regular at $q^2=0$.
The second equation in (\ref{constraint-polarization})
implies the following  constraint,
\begin{eqnarray}
&&\Pi_3(\omega, 0) =\frac{1}{\omega^2}\,\Pi_1(\omega,0) \,, \qquad \qquad \quad
\Pi_2(\omega, 0)= \frac{1}{\omega}\,\Pi_1(\omega,0)  \,,
\nonumber\\
&& \Pi_4(\omega,0) = 3\,\Pi_{22}(\omega,0)+\Pi_{11}(\omega, 0)\,,
\end{eqnarray}
where we boosted into the rest frame of nuclear matter for convenience.
Based on the representation (\ref{S-decompose}) we define
\begin{eqnarray}
&& \Pi_{i}^{(\Delta h)} (\omega , \vec q\,)= \Bigg[\delta_{i4}\,\Pi^{(\Delta h)}_3(0, \vec q\,)
\nonumber\\
&& \qquad -
\frac{8}{3}\,\frac{f_\Delta^2}{m_\pi^2}
\int_0^{k_F}\frac{d^3p}{2\,(p_0-\Sigma_V)\,(2\pi)^3} \int_{-\infty }^{+\infty } \frac{d \bar \omega }
{\pi}\,\left( \frac{\omega}{\bar \omega}\right)^{n_i}\,
\frac{\sign (\bar \omega)\,\Im S^{(\Delta h)}_{i} (\bar \omega , \vec q, \vec p\,) }
{\bar \omega - \omega- i\,\bar \omega \,\epsilon} \Bigg]
\nonumber\\
&& \qquad  \;+(-1)^{\epsilon_i} (q_\mu \to -q_\mu )\,,
\label{dh-int-k}
\end{eqnarray}
where $p_0 =\sqrt{m_N^2+\vec p\,^2}+\Sigma_V$ and $\epsilon_{1,3,4}=0$
and $\epsilon_2 =1$. Furthermore $n_{1,4}=2$ but $n_{2}=1$ and $n_3=0$. We assure that the definition
(\ref{dh-int-k}) leads to a polarization tensor compatible with all
constraints (\ref{constraint-polarization}, \ref{bounded-assumption}).
This is a consequence of specific identities the integral kernels enjoy (see \ref{2nd-condition-rewrite-s}).

The integral kernels, $S^{(\Delta h)}_{i}(q,p,u)$, required in (\ref{dh-int-k})
are covariant functions of the  4-momenta $q_\mu, p_\mu$ and $u_\mu$. Their evaluation
requires the contraction of the isobar propagator, $S_{\mu \nu}(p+q,u)$, with
the $q_\mu$ and $u_\mu$ (see (\ref{nh-dh}, \ref{add22b})).
We express the 4-vector $u_\mu$, in terms of $v_\mu$
and $X_\mu (v,u)$,
\begin{eqnarray}
u_\mu = -\sqrt{(\hat v\cdot u)^2-1}\,X_\mu (v,u) + (\hat v \cdot u)\,\hat v_\mu \,,
\label{ux}
\end{eqnarray}
since the contraction of the isobar propagator with $v_\mu$
and $X_\mu(v,u)$ leads to more transparent expressions. In particular we can take over
the results from \cite{Lutz:Korpa:Moeller:2007}, where contractions of the isobar
propagator with the latter 4-vectors were computed already. The results were decomposed into
the extended algebra of projectors (\ref{q-space-def}, \ref{p-space-def})
introducing the invariant expansion coefficients $S^{(a)}_{[ij]}(v,u)$ and $S^{(ab)}_{[ij]}(v,u)$
with $a,b = v,x$.

We present the integral kernels of (\ref{dh-int-k}), which have
transparent representations in terms of the invariant functions introduced in
(\ref{S-decompose})
and $c^{(p,q)}_{[ij]}(q;v,u)$ of \cite{Lutz:Korpa:Moeller:2007}. We establish:
\begin{eqnarray}
&& S^{(\Delta h)}_{1} = \sum_{i,j=3}^8 \,c_{[ij]}^{(p)}\,S^{(p)}_{[ij]}
+\sum_{i,j=1}^2 \,c_{[ij]}^{(q)}\,S^{(q)}_{[ij]} \,,
\nonumber\\
&& S^{(\Delta h)}_{2} = \sum_{i=1}^2\,\sum_{j=3}^8\, c^{(p)}_{[ij]} \,\Big[
(\hat v \cdot u)\,S^{(v)}_{[ij]}
- \sqrt{(\hat v \cdot u)^2-1}\,S^{(x)}_{[ij]}\Big] \,,
\nonumber\\
&& S^{(\Delta h)}_{3} = \sum_{i,j=1}^2\,c^{(p)}_{[ij]}
\Big[ (\hat v \cdot u)^2\,S^{(vv)}_{[ij]}
+ \Big((\hat v \cdot u)^2-1 \Big)\,
S^{(xx)}_{[ij]}
\nonumber\\
&& \qquad \qquad \qquad \qquad \qquad - (\hat v \cdot u)\,
\sqrt{(\hat v \cdot u)^2-1}\,\Big( S^{(xv)}_{[ij]}
+S^{(vx)}_{[ij]}  \Big)
\Big] \,,
\nonumber\\
&&S^{(\Delta h)}_{4} = \sum_{i,j=1}^2\, c_{[ij]}^{(p)}\,S^{(g)}_{[ij]}\,.
\label{Sdh-final}
\end{eqnarray}
A straight forward computation reveals that the kernels $S^{(\Delta h)}_i$ are correlated at
vanishing 3-momentum $\vec q\,=0$. In this case it holds
\begin{eqnarray}
&& S^{(\Delta h)}_3 =\frac{1}{\omega^2}\,S^{(\Delta h)}_1 \,, \qquad \qquad S^{(\Delta h)}_2=
\frac{1}{\omega}\,S^{(\Delta h)}_1  \,,
\label{2nd-condition-rewrite-s}\\
&& S^{(\Delta h)}_4 = 3\,S_3^{(\Delta h)}- \frac{2}{\omega^2}\,S_1^{(\Delta h)}
-3\,\frac{d}{d \vec q\,^2}\Bigg|_{\vec q=0}\,
\Big( S_1^{(\Delta h)}-2\,\omega\,S_2^{(\Delta h)}+\omega^2\,S_3^{(\Delta h)}
\Big)\,, \nonumber
\end{eqnarray}
where we assumed an angle average, i.e. the presence of $d \Omega_{\vec q}$.

\section*{Appendix B}

We derive

\begin{eqnarray}
&&V^{(p)}_{[33]}=\frac{f_{\Delta}^2}{m_{\pi}^2}\left[\frac{2\,\delta_V\,\delta \,(m_{\Delta} + \wws)}{9\, m_{\Delta}^2 (m_{\Delta}^2 - \w2)} \right],\nonumber\\[1.3ex]
&&V^{(p)}_{[34]}=\frac{i\,\delta\,f_{\Delta}^2}{m_{\pi}^2\,m_{\Delta}^2 \sqrt{\uws^2 - 1} }\left[\frac{-2\,\delta_V\,\delta }{9\,(m_{\Delta}^2 - \w2)}
                 \right],\nonumber\\[1.3ex]
&&V^{(p)}_{[35]}=\frac{f_{\Delta}^2}{\sqrt{3}\,m_{\pi}^2\,m_{\Delta}^2}\left[\frac{3\, m_{\Delta}\,(m_{\Delta}^2-\wws^2)
        -\delta_V\,\delta\,(2\,\wws-m_{\Delta})}{3\,  (m_{\Delta}^2 - \w2) }\right],\nonumber\\[1.3ex]
&&V^{(p)}_{[36]}=-V^{(p)}_{[45]}=\frac{i\,\delta\,f_{\Delta}^2}{m_{\pi}^2\,m_{\Delta}^2\, \sqrt{3\, \uws^2-3} }\left[\frac{2\,
                 (m_{\Delta}^2  - \wws^2)}{3\, (m_{\Delta}^2 - \w2) } \right],\nonumber\\[1.3ex]
&&V^{(p)}_{[37]}=\frac{\sqrt{2}\,i\,\delta\, f_{\Delta}^2}{3\, m_{\pi}^2\,m_{\Delta}^2\,\sqrt{\uws^2 - 1}}\left[\frac{3\,
                 m_{\Delta}(m_{\Delta} + \wws) + 2\,\delta_V\,\delta }{3\,(m_{\Delta}^2 - \w2)} \right],\nonumber\\[1.3ex]
&&V^{(p)}_{[38]}=\frac{-f_{\Delta}^2\,\delta\, (\delta_V\, ( \ws2-\uws^2)+3\, \delta)\, (m_{\Delta} - 2\, \wws)}{
                 9\, \sqrt{2}\, m_{\pi}^2\,m_{\Delta}^2\, (m_{\Delta}^2 - \w2)\, ( \ws2-\uws^2)},\nonumber\\[1.3ex]
&&V^{(p)}_{[44]}=\frac{f_{\Delta}^2}{m_{\pi}^2\,m_{\Delta}^2}\left[\frac{2\, \delta_V\,\delta \,( \wws-m_{\Delta})}{9\,  (m_{\Delta}^2 - \w2)
                 } \right],\nonumber\\[1.3ex]
&&V^{(p)}_{[46]}=\frac{f_{\Delta}^2}{\sqrt{3}\, m_{\pi}^2\,m_{\Delta}^2 }\left[\frac{-3\, m_{\Delta}\,(m_{\Delta}^2-\wws^2)
        -\delta_V\,\delta\,(2\,\wws+m_{\Delta})}{3\, (m_{\Delta}^2 - \w2)}\right],\nonumber\\[1.3ex]
&&V^{(p)}_{[47]}=\frac{-f_{\Delta}^2\,\delta\, (\delta_V\, ( \ws2-\uws^2)+3\, \delta)\, (m_{\Delta} + 2\, \wws)}{
                 9\, \sqrt{2}\, m_{\pi}^2\,m_{\Delta}^2\, (m_{\Delta}^2 - \w2)\, ( \ws2-\uws^2)},\nonumber\\[1.3ex]
&&V^{(p)}_{[48]}=\frac{\sqrt{2}\,i\,\delta\, f_{\Delta}^2}{3\, m_{\pi}^2\,m_{\Delta}^2\,\sqrt{\uws^2 - 1}}\left[\frac{3\,
                 m_{\Delta}(m_{\Delta} - \wws) + 2\,\delta_V\,\delta }{3\,(m_{\Delta}^2 - \w2)} \right],\nonumber\\[1.3ex]
&&V^{(p)}_{[55]}=\frac{f_{\Delta}^2}{3\,m_{\pi}^2\, m_{\Delta}^2 }\left[\frac{-2\, (m_{\Delta} - \wws)\, (m_{\Delta} + \wws)^2}{
                  m_{\Delta}^2 - \w2}\right],\nonumber\\[1.3ex]
&&V^{(p)}_{[56]}=\frac{i\,\delta\,f_{\Delta}^2}{m_{\pi}^2 \,m_{\Delta}^2\, \sqrt{\uws^2 - 1}}\left[\frac{-2\,(m_{\Delta}^2 -
                  \wws^2)}{3\, (m_{\Delta}^2 - \w2) } \right],\nonumber\\[1.3ex]
&&V^{(p)}_{[57]}=-\frac{\sqrt{2}\,i\,\delta\, f_{\Delta}^2}{3\,\sqrt{3}\, m_{\pi}^2\,m_{\Delta}^2\,\sqrt{\uws^2 - 1}}\left[\frac{
                 m_{\Delta}^2 + 3\, m_{\Delta} \, \wws + 2\,\wws^2  }{(m_{\Delta}^2 - \w2)} \right],\nonumber\\[1.3ex]
&&V^{(p)}_{[58]}=\frac{f_{\Delta}^2\,\delta\, (\delta_V\, ( \ws2-\uws^2)+3\, \delta)\, (m_{\Delta} - 2\, \wws)}{
                  3\, \sqrt{6}\, m_{\pi}^2\,m_{\Delta}^2\, (m_{\Delta}^2 - \w2)\, ( \ws2-\uws^2)},\nonumber\\[1.3ex]
&&V^{(p)}_{[66]}=\frac{f_{\Delta}^2}{3\,m_{\pi}^2\, m_{\Delta}^2 }\left[\frac{-2\, (m_{\Delta} - \wws)^2\, (m_{\Delta} + \wws)}{
                 m_{\Delta}^2 - \w2}\right],\nonumber\\[1.3ex]
&&V^{(p)}_{[67]}=-\frac{f_{\Delta}^2\,\delta\, (\delta_V\, ( \ws2-\uws^2)+3\, \delta)\, (m_{\Delta} + 2\, \wws)}{
                  3\, \sqrt{6}\, m_{\pi}^2\,m_{\Delta}^2\, (m_{\Delta}^2 - \w2)\, ( \ws2-\uws^2)},\nonumber\\[1.3ex]
&&V^{(p)}_{[68]}=\frac{\sqrt{2}\,i\,\delta\, f_{\Delta}^2}{3\,\sqrt{3}\, m_{\pi}^2\,m_{\Delta}^2\,\sqrt{\uws^2 - 1}}\left[\frac{
                  m_{\Delta}^2 - 3\, m_{\Delta} \, \wws + 2\,\wws^2  }{(m_{\Delta}^2 - \w2)} \right],\nonumber\\[1.3ex]
&&V^{(p)}_{[77]}=-\frac{f_{\Delta}^2 \,(m_{\Delta} + \wws)}{ m_{\pi}^2\, (m_{\Delta}^2 -
                  \w2)}\nonumber\\[1.3ex]
                  &&\qquad+\frac{f_{\Delta}^2\,\delta\,(3\, \delta\, (2\, m_{\Delta} + \wws) +\delta_V\,(\uws^2 - \ws2)\, (2\,m_{\Delta} + \wws))}{
                  9\, m_{\pi}^2\, m_{\Delta}^2\, (m_{\Delta}^2 - \w2)\, ( \ws2-\uws^2)},\nonumber\\[1.3ex]
&&V^{(p)}_{[78]}=\frac{-i\, \delta\, f_{\Delta}^2\, (9\, \delta^2 + (\uws^2 - \ws2)\, (-5\,\delta_V\,\delta + 3\, m_{\Delta}^2))}{9\,
                  m_{\pi}^2 \,m_{\Delta}^2\, (m_{\Delta}^2 - \w2) \,\sqrt{\uws^2 - 1}^3 },\nonumber\\[1.3ex]
&&V^{(p)}_{[88]}=-\frac{f_{\Delta}^2 \,(m_{\Delta} - \wws)}{ m_{\pi}^2\, (m_{\Delta}^2 -
                 \w2)}\nonumber\\[1.3ex]
                 &&\qquad+\frac{f_{\Delta}^2\,\delta\,(3\, \delta\, (2\, m_{\Delta} - \wws) +\delta_V\,(\uws^2 - \ws2)\,
                 (2 \,m_{\Delta} - \wws))}{
                 9\, m_{\pi}^2\, m_{\Delta}^2\, (m_{\Delta}^2 - \w2)\, ( \ws2-\uws^2)},\nonumber\\[1.3ex]
&&V^{(q)}_{[11]}=\frac{f_{\Delta}^2\,(m_{\Delta}+\wws)}{m_{\pi}^2\,(\w2-m_{\Delta}^2)},\qquad V^{(q)}_{[22]}=\frac{f_{\Delta}^2\,(m_{\Delta}-\wws)}{m_{\pi}^2\,(\w2-m_{\Delta}^2)},\nonumber\\[1.3ex]
&&V^{(q)}_{[12]}=\frac{i\,\delta\,f_{\Delta}^2}{m_{\pi}^2\,(\w2-m_{\Delta}^2)\,\sqrt{\uws^2-1}},
\end{eqnarray}
%
%
where
\begin{eqnarray}
&&\tilde w_\mu = w_\mu- \Sigma^\Delta_V\,u_\mu\,,
\quad \delta=\uws\wws-\uw\,, \quad \delta_V = \Sigma^N_V-\Sigma^\Delta_V\,.
\end{eqnarray}


\newpage

\begin{thebibliography}{9}
\bibitem{Campbell}
D. Campell, R. Dashen and J. Manassash, Phys. Rev. {\bf D 12} (1975) 979.
\bibitem{Oset:Weise:76}
E. Oset and W. Weise, Phys. Lett. {\bf B 60} (1976) 141.
\bibitem{Migdal:1978}
A.B. Migdal, Rev. Mod. Phys. {\bf 50} (1978) 107.
\bibitem{Oset:Weise}
E. Oset, H. Toki and W. Weise, Phys. Rep. {\bf 83} (1982) 281.
\bibitem{Dyugaev}
A.M. Dyugaev, Sov. J. Nucl. Phys. {\bf 38} (1983) 680.
\bibitem{Dmitriev:Suzuki}
V.F. Dmitriev and T. Suzuki, Nucl. Phys. {\bf A 438} (1985) 697.
\bibitem{Oset:Salcedo}
E. Oset and L.L. Salcedo, Nucl. Phys. {\bf A 468} (1987) 631.
\bibitem{Migdal}
A.B. Migdal et al.,
Phys. Rep. {\bf 192} (1990) 181.
\bibitem{Herbert:Wehrberger:Beck}
T. Herbert, K. Wehrberger and F. Beck, Nucl. Phys. {\bf A 541} (1992) 699.
\bibitem{Carrasco:Oset}
R.C. Carrasco and E. Oset, Nucl. Phys. {\bf A 536} (1992) 445.
\bibitem{Nieves:Oset:Recio}
J. Nieves, E. Oset and C. Garcia-Recio, Nucl. Phys. {\bf A 554}
(1993) 554.
\bibitem{Xia:Siemens:Soyeur}
L. Xia, P.J. Siemens and M. Soyeur, Nucl. Phys. {\bf A 578} (1994) 493.
\bibitem{Arve:Helgesson}
P. Arve and J. Helgesson, Nucl. Phys. {\bf A 572} (1994) 600.
\bibitem{Korpa:Malfliet}
C.L. Korpa and R. Malfliet, Phys. Rev. {\bf C 52} (1995) 2756.
\bibitem{Schramm}
H. Kim, S. Schramm and S.H. Lee, Phys. Rev. {\bf C 56} (1997) 1582.
\bibitem{Rapp}
R. Rapp, M. Urban, M. Buballa and J. Wambach. Phys. Lett. {\bf B 417} (1998) 1.
\bibitem{Nakano}
M. Nakano et al., Int. J. Mod. Phys. {\bf E 10} (2001) 459.
\bibitem{Lutz:Migdal}
M.F.M. Lutz, Phys. Lett. {\bf B 552} (2003) 159; Erratum ibd {\bf
B 566} (2003) 277.
\bibitem{Korpa:Lutz:04}
C.L. Korpa and M.F.M. Lutz, Nucl. Phys. {\bf A 742} (2004) 305.
\bibitem{Post:Leupold:Mosel}
M. Post, S. Leupold and U. Mosel, Nucl. Phys. {\bf A 741} (2004) 81.
\bibitem{Knoll}
F. Riek and J. Knoll, Nucl. Phys. {\bf A 740} (2004) 287.
\bibitem{KoDi04}
C.L. Korpa and A.E.L. Dieperink, Phys. Rev. {\bf C 70} (2004) 015207.
\bibitem{Hees:Rapp}
H. van Hees and R. Rapp, Phys. Lett. {\bf B 606} (2005) 59.
\bibitem{Lutz:Kolomeitsev}
M.F.M. Lutz and E.E. Kolomeitsev, Nucl. Phys. {\bf A 700} (2002) 193.
\bibitem{Lutz:Korpa:02}
M.F.M. Lutz and C.L. Korpa, Nucl. Phys. {\bf A 700} (2002) 309.
\bibitem{Lutz:Korpa:Moeller:2007}
M.F.M. Lutz, C.L. Korpa  and M. M\"oller, Nucl. Phys. {\bf A 808} (2008) 124.
\bibitem{Riek:Lutz:Korpa:2008}
F. Riek, M.F.M. Lutz and  C.L. Korpa, arXiv:0809.4608 [nucl-th].
\bibitem{Hirata:Koch:Lenz:Monitz}
M. Hirata, J.H. Koch, F. Lenz and E.J. Moniz, Ann. Phys. {\bf 120} (1979) 205.
\bibitem{photo-absorption}
J. Ahrens et al., Phys. Lett, {\bf B 146} (1984) 303; N. Bianchi
et al., Phys. Lett. {\bf B 299} (1993) 219; Th. Frommhold et al.,
Zeit. Phys. {\bf A 350} (1994) 249; N. Bianchi et al., Phys. Rev.
{\bf C 54} (1996) 1688.
\bibitem{Wakasa}
T. Wakasa et al., Phys. Rev. {\bf C 55} (1997) 2909.
\bibitem{Arndt}
SAID on-line programm, http://gwdac.phys.gwu.edu/.
\bibitem{Ramos:Oset:2000}
A. Ramos and E. Oset, Nucl. Phys. {\bf A 671} (2000) 481.
\end{thebibliography}
\end{document}